%% file: cap_jrs9.TEX
\documentclass[12pt,draftclsnofoot,onecolumn]{IEEEtran} 

\usepackage{times}
\usepackage[final]{graphicx}
\usepackage{amsmath,amsfonts}
\usepackage{amssymb,amsbsy}
\usepackage{times}

\usepackage{epsf,graphics,epsfig,subfigure} 
\usepackage{cite}
\usepackage{multirow}

\input{macros}

\newcommand{\bal}{\begin{align}}
\newcommand{\eal}{\end{align}}

\newcommand{\bP}{{\bf P}}
\newcommand{\bhu}{ {{\bf h}}}

\newsavebox{\savepar}

\begin{document}
\title{Why Does the Kronecker Model Result in Misleading Capacity Estimates?} 
\author{
\large Vasanthan Raghavan$^*$, Jayesh H. Kotecha, and Akbar M. Sayeed 
\thanks{V. Raghavan is with the Coordinated Science Laboratory and the Department 
of Electrical and Computer Engineering, University of Illinois at Urbana-Champaign, 
Urbana, IL 61801 USA. J. H. Kotecha is with Freescale Semiconductor Inc., Austin, 
TX 78721 USA, and A. M. Sayeed is with the Department of Electrical and Computer 
Engineering, University of Wisconsin-Madison, Madison, WI 53706 USA. 
Email: {\sf{vasanthan\_\_raghavan@ieee.org, jayeshkotecha@freescale.com, 
akbar@engr.wisc.edu.}} 
$^*$Corresponding author. 
This research was supported in part by NSF Grant \#CCF-0431088 
through the University of Wisconsin.}} 

\maketitle
\vspace{-10mm}
\baselineskip 17pt

\begin{abstract} 
\noindent 
Many recent works that study the performance of multi-input multi-output 
(MIMO) systems in practice assume a Kronecker model where the variances 
of the channel entries, upon decomposition on to the transmit and the receive 
eigen-bases, admit a separable form. Measurement campaigns, however, show 
that the Kronecker model results in poor estimates for capacity. Motivated 
by these observations, a channel model that does not impose a separable 
structure has been recently proposed and shown to fit the capacity of 
measured channels better. In this work, we show that this recently proposed 
modeling framework can be viewed as a natural consequence of channel 
decomposition on to its {\em canonical coordinates}, the transmit and$/$or 
the receive eigen-bases. Using tools from random matrix theory, we then 
establish the theoretical basis behind the Kronecker mismatch at the low- 
and the high-$\snr$ extremes: 1) {\em Sparsity} of the dominant statistical 
degrees of freedom (DoF) in the true channel at the low-$\snr$ extreme, 
and 2) {\em Non-regularity} of the sparsity structure (disparities in the 
distribution of the DoF across the rows and the columns) at the high-$\snr$ 
extreme. 
\end{abstract}


\section{Introduction} 
Under the assumption of spatially independent and identically distributed 
(i.i.d.) Rayleigh fading between antenna pairs, multi-input multi-output 
(MIMO) systems achieve a linear growth in multiplexing gain and coherent 
capacity with the number of antennas~\cite{Telatar,Foschini}. However, 
the rich scattering assumption is idealistic and most physical channels 
encountered in practice exhibit clustered scattering and spatially 
correlated links~\cite{Chuah,Shiu,review_bol}. Correlated MIMO channels 
have been theoretically studied mainly in the contexts of the separable 
correlation model (also known as the {\em Kronecker model})~\cite{Chuah,Shiu}, 
and the virtual representation framework for uniform linear arrays 
(ULAs)~\cite{Sayeeddecon,spl_issue,Veeravallicap,ada}. The Kronecker model 
assumes separability in correlation induced by the transmitter and the 
receiver arrays which limits the degrees of freedom (DoF) in modeling the 
channel. Though this model has been shown to be accurate in certain 
settings (especially $2 \times 2$ 
scenarios)~\cite{Kai,mcnamara,kron_int1,wallace,mimo_manhattan}, 
the separability assumption limits its applicability to more realistic 
settings where the gains accrued with MIMO make it a viable choice. 
The virtual representation does not assume such separability, but is 
applicable only for ULAs. 

\noindent {\bf \em Contributions}: 
\begin{itemize} 
\item 
i) In this paper, we develop a unified statistical modeling framework for 
Rayleigh fading MIMO channels based on decomposition of the channel matrix 
on its {\em canonical coordinates}, the transmit and$/$or the receive 
eigen-bases. Motivated by virtual representation, these models do not 
assume separable statistics and are applicable to general array 
geometries. Like the Kronecker model, the eigen-modes of the scattering 
environment decide the transmit and the receive eigen-bases whereas the 
canonical channel matrix embodies the statistically independent DoF that 
govern channel capacity and diversity. 

Depending on the covariance structure of the channel, three models arise 
in which all the columns$/$rows$/$channel entries are uncorrelated. The 
last case\footnote{Some of the earlier works of the 
authors~\cite{KotechachannelestICC,Kotechachannelest} have also suggested 
this model and developed it independently~\cite{Kotechacaptech}.}, denoted 
here as the {\em canonical model} (or $\CM$ for short), has been proposed as 
the {\em Weichselberger model} in~\cite{Bonek} and studied from a capacity 
analysis viewpoint in~\cite{tulino_ind}. 
The new contribution in this work is the unified development of $\CM$ 
as a natural consequence of two other models, denoted as $\CMone$ and 
$\CMtwo$. The development of $\CMone$ (and $\CMtwo$) critically depends 
on two assumptions about the covariance and the cross-covariance information 
of the rows (and the columns) of the channel matrix. We establish the 
criticality of these four assumptions in the development of $\CM$. To the 
best of our knowledge, $\CMone$ and $\CMtwo$ have not been proposed 
elsewhere in the literature, and could provide useful intermediate models 
for certain asymmetric MIMO systems.

\item 
ii) Many recent 
works~\cite{Bonek,zhou,costa_haykin,new_ozcelik,wyne,abhayapala,survey,bonek_valid} 
have shown that the Kronecker model consistently estimates the capacity of 
a large class of measured channels poorly and hence they establish 
the need\footnote{In fact, the development of $\CM$ in~\cite{Bonek} is 
motivated by these observations.} for more accurate channel modeling. 
For example,~\cite{Bonek,zhou,costa_haykin,new_ozcelik} show that the 
Kronecker model severely underestimates true capacity whereas under 
certain conditions, it could also overestimate the true 
capacity~\cite{abhayapala,bonek_valid}. 
Nevertheless, motivated by extensive measurement studies, the popular belief 
is that the ``probability of overestimation decreases with increasing antenna 
number''~\cite[Footnote 5]{bonek_valid}. The main focus of this work is to 
theoretically explain these observations. 

Towards this goal, we first note that recent measurement campaigns 
have also observed that only a few of the statistical DoF are dominant 
enough to contribute towards reliable communications. That is, measured 
multi-antenna channels are {\em sparse} in the canonical domain. 
Furthermore, the distribution 
of the sparse DoF across the spatial domain does not observe any 
{\em regularity}\footnote{\label{footnote_regular}Let $\bH_c$ be an 
$N_r \times N_t$ random matrix with independent entries and let the 
variance of $\bH_c[i,j]$ be given by $\cvarij$. A channel is called 
{\em column-regular} if $\sum_{i=1}^{N_r} \cvarij$ is equal for 
all $j$, {\em row-regular} if the above condition is true for $\bH_c^T$, 
and {\em regular} if it is both row- and column-regular~\cite{Tulino}. 
Otherwise, it is {\em non-regular.}} structure. For example, 
see~\cite[Figs.\ 9 and 11]{bonek_valid},~\cite{wood_hodgkiss,costa_haykin,zhou} 
etc.\ which plot the sparse, non-regular structure of the 
{\em Weichselberger coupling matrix} that reflects the statistical DoF in 
$\CM$.

{\vspace{-0.2in}}
\begin{figure}[h]
\centerline{\psfig{figure=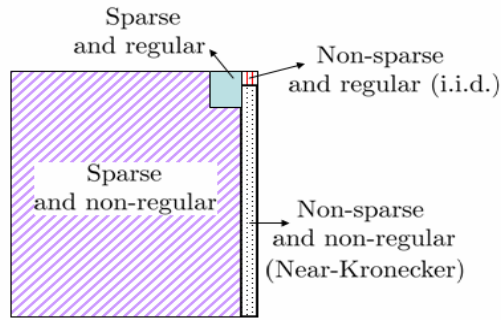,width=5.0in,height=3.4in}}
{\vspace{-1.2in}}
\caption{Partitioning of the space of all possible channels based 
on sparsity and regularity.}
\label{fig_sparse}
{\vspace{-0.1in}}
\end{figure}

The foundation for the above experimental evidence lies in theoretical 
electromagnetic studies that explain sparsity of DoF in different contexts 
in wireless 
communications~\cite{bucci,migliore,ada,massimo,xu_janaswamy,hanlen_sparse,rod_kennedy_sparse,goodman}. 
Nevertheless, a simpler communication-theoretic motivation for sparsity is 
that while there may be many channel coefficients whose energy levels are 
non-zero, they may not be strong enough to be estimated accurately at the 
transmitter, even statistically. It becomes impossible or too costly to 
estimate such coefficients accurately and thus from the transmitter's 
viewpoint, it is reasonable to treat their contributions as noise. Thus, 
we can partition the space of all possible channels into four classes, as 
in Fig.~\ref{fig_sparse}, with the class of sparse and non-regular channels 
being the most predominant. 
In this work, we develop a mathematical framework for probabilistically 
modeling sparse multi-antenna channels. The framework developed here 
allows us to adjust the average number of dominant channel coefficients 
and theoretically study the impact of a Kronecker model on the capacity 
mismatch. 

\item 
iii) We assume that the statistics of the true channel in the canonical 
domain has a sparse, non-separable structure. 
Based on recent works~(see~\cite{it_rs01} and references therein) that 
establish the accuracy of a Gaussian approximation to outage capacity, 
we also assume that the ergodic capacity ($C_{\erg}(\rho)$) and the 
variance of capacity ($V(\rho)$) are the key figures-of-merit. The main 
results of this work are obtained for the low- and the high-$\snr$ 
extremes in the large-system (antenna) regime, and are summarized in 
Table~\ref{table1}.

\begin{center}
\label{table1} 
\begin{tabular}{|c|c|c|}
\hline 
\hline 
\multicolumn{3}{|c|}{Table I: Summary of Main Results} \\ 
\hline 
\hline 
& Mismatch Metric & Conclusions \\ \hline 
\multirow{4}{*}{$\rho \rightarrow 0$} 
& $C_{\erg,\hspp\can}(\rho) = C_{\erg,\hspp \kron}(\rho)$ for 
all chan.\ & \\ 
& $\frac{V_{\can}(\rho)}{V_{\kron}(\rho)} \rightarrow 
\frac{1}{p} \left(1 + \frac{\sigma^2}{\mu^2} \right)$ where 
& I.I.D.\ $\Longleftarrow {\bf H}_c \Longrightarrow$ Sparse 
\\ 
& $\{p, \mu, \sigma \}$ are sparse model parameters & 
Decreases $\Longleftarrow$ Mism.\ $\Longrightarrow$ Increases \\ 
& (See Theorem 2 and Prop.~2 for details) & 
\\ 
\hline 
\multirow{6}{*}{$\rho \rightarrow \infty$} 
& $C_{\erg,\hspp\can}(\rho) - C_{\erg,\hspp\kron}(\rho) \rightarrow$
&  
\\ 
& $\frac{N}{2} \log_2 \left( \frac{ {\rm AM}_{\sf row \hspp pow} 
\cdot {\rm AM}_{\sf col \hspp pow} }  
{ {\rm GM}_{\sf row \hspp pow} \cdot {\rm GM}_{\sf col \hspp pow}  }
\right)$ & 
\\ 
& where ${\rm AM}_{\bullet}$ and ${\rm GM}_{\bullet}$ are 
arithmetic & 
Regular $\Longleftarrow {\bf H}_c \Longrightarrow$ Non-regular 
\\  
& and geometric means of row and & 
Decreases $\Longleftarrow$ Mism.\ $\Longrightarrow$ Increases
\\
& column powers of the true channel & \\
& (See Theorem 4 for details) & \\ 
\hline 
\hline 
\end{tabular}
\end{center}

{\vspace{0.1in}}
We show that for {\em almost every} sparse channel: a) Using marginal 
sum statistics to generate a Kronecker fit results in an artificial increase 
in the number of DoF, b) As a result, the channel power is spread across 
the increased DoF, and c) Hence, the Kronecker model offers a poor estimate 
for capacity. Towards establishing this connection, we develop a tight 
approximation for the mean of the log-determinant of random matrices (with 
independent entries) which is of independent interest in MIMO analysis and 
design. 

In the high-$\snr$ extreme, the Kronecker model {\em underestimates} 
$C_{\erg}(\rho)$ for all channels. The level of underestimation decreases 
as the channel becomes more regular. In the low-$\snr$ extreme, 
$C_{\erg}(\rho)$ is the same with either model. The Kronecker model 
underestimates $V(\rho)$ with the level of underestimation decreasing 
as the channel becomes less sparse. Thus, for a large class of channels 
that are sparse and non-regular, the Kronecker model {\em underestimates} 
the outage capacity at all reliability levels (and also the reliability at 
all data rates) in the medium- to high-$\snr$ regime. On the other hand, 
for any channel in the low-$\snr$ regime, and regular channels in the 
medium- to high-$\snr$ regime, the Kronecker model {\em overestimates} 
capacity at high levels of reliability (and reliability at low data rates), 
and {\em vice versa}. 
\end{itemize}

\noindent {\bf \em Organization}: 
This paper is organized as follows. The canonical statistical modeling 
framework for correlated multi-antenna channels is developed in 
Section~\ref{MIMO} with the key properties of the proposed model 
elucidated in Section~\ref{cm3_features}. In Section~\ref{sec_prac}, we 
explore practical modeling issues and show how the canonical and the 
Kronecker models are used to describe realistic measured channels. A brief 
summary of MIMO capacity issues is provided in Section~\ref{coh_cap_norm} 
with a comparative study of the two 
models performed 
in Section~\ref{capacity}. Conclusions are drawn in Section~\ref{conclusion}. 

\noindent {\bf \em Notation}: 
We use upper-case and lower-case bold symbols for matrices and vectors, respectively. 
If $\bX$ is an $M \times N$ matrix, ${\bf x} = {\rm vec}(\bX)$ denotes the $MN 
\times 1$ vector obtained by stacking columns of $\bX $. The entry in the $m$-th 
row and $n$-th column, and the $m$-th diagonal entry of $\bX$ are denoted by 
$\bX[m,n]$ and $\bX[m] = \bX[m,m]$, respectively. The complex conjugate, regular 
transpose and Hermitian transpose of $\bX$ are denoted by $\bX^*, \bX^T$ and $\bX^H$ 
while its inverse, trace and determinant are denoted by $\bX^{-1}$, ${\rm Tr}(\bX)$ 
and ${\rm det}(\bX)$, respectively. The operators $E[\cdot]$, $\otimes$ 
and $\odot$ stand for expectation, Kronecker and Hadamard products. The indicator 
function of a set ${\mathcal{A}}$ and its probability are given by 
$\chi( {\mathcal{A}})$ and ${\mathrm {Pr}}({\mathcal{A}})$. We use the 
standard big-Oh ($\ord$) and little-oh ($\littleo$) notations, $\sim$ for 
equality in distribution, and 
$X \sim \compnorm(\mu,\sigma^2)$ to indicate that $X$ is a complex Gaussian 
random variable with mean $\mu$ and variance $\sigma^2$.

\section{Canonical Modeling of Correlated MIMO Channels}
\label{MIMO} 
Consider a narrowband, Rayleigh fading MIMO channel with $N_t$ 
transmit and $N_r$ receive antennas. The $N_r \times 1$ received 
vector ${\bf y}$ is related to the $N_t \times 1$ transmit vector 
${\bf x}$ by 
\beq
\label{rec}
\by = {\bf H}{\bf x}  + {\bf n}
\eeq
where $\bH$ is the $N_r \times N_t$ channel matrix and ${\bf n}$ is 
the independent, white Gaussian noise added at the receiver. The entries 
of $\bH$ are zero mean, complex Gaussian that satisfy 
\begin{eqnarray} 
\bh \triangleq {\rm vec}(\bH) \sim \compnorm({\bf 0}, {\bf R})
\end{eqnarray} 
for some positive semi-definite channel covariance matrix ${\bf R}$. 

We now describe three canonical decompositions of MIMO channels. Let 
$\bQi_t \triangleq E \left[ \bH^H \bH \right]$ and 
$\bQi_r \triangleq E \left[ \bH \bH^H \right]$ denote the transmit and the 
receive covariance matrices. Let their respective eigen-decompositions be given 
by $\bQi_t = \bU_t \bLambda_t \bU_t^H$ and $\bQi_r = \bU_r \bLambda_r \bU_r^H$ where 
the columns of $\bU_t$ and $\bU_r$ are eigenvectors of $\bQi_t$ and $\bQi_r$ with 
the corresponding eigenvalues denoted by diagonal entries of $\bLambda_t$ and 
$\bLambda_r$. 
\begin{lemma} 
Any channel matrix $\bH$ can be written in the canonical form: 
$\bH = \bH_t \bU_t^H$ such that $E \left[\bH_t^H \bH_t \right] = \bLambda_t$ 
and 
\begin{eqnarray}
\bh_t & \triangleq & {\rm vec}(\bH_t) = {\rm vec}(\bH \bU_t) = 
(\bU_t^T \otimes \bI) \hsppp{\bf h} \sim \compnorm({\bf 0}, \bR_t) \\ 
\bR_t & \triangleq & E \left[\bh_t \bh^H_t \right] 
=(\bU_t^T \otimes \bI) \hspp \bR  \hspp (\bU_t^T \otimes \bI)^H. 
\end{eqnarray} 
Similarly, any channel matrix $\bH$ can be written in the canonical form: 
$\bH = \bU_r \bH_r$ such that $E \left[\bH_r \bH_r^H \right] = \bLambda_r$ and 
\begin{eqnarray} 
\bh_r & \triangleq & 
{\rm vec}(\bH_r) \sim \compnorm({\bf 0}, \bR_r) \\ 
\bR_r & \triangleq & E \left[\bh_r \bh^H_r \right] = 
(\bI \otimes \bU_r^H) \hspp \bR \hspp (\bI \otimes \bU_r). 
\end{eqnarray} 
Furthermore, $\bH$ can also be written in the canonical form: 
$\bH = \bU_r \bH_c \bU_t^H$ such that $E \left[\bH_c^H \bH_c \right] 
= \bLambda_t$, $E \left[\bH_c \bH_c^H \right] = \bLambda_r$ and 
\begin{eqnarray}
\bh_c & \triangleq & {\rm vec}(\bH_c) \sim \compnorm({\bf 0}, \bR_c) \\ 
\bR_c & \triangleq & E \left[ \bh_c \bh^H_c \right] = 
(\bU_t^T \otimes \bU_r^H) \hspp \bR \hspp (\bU_t^T \otimes \bU_r^H)^H. 
\end{eqnarray} 
\end{lemma}
\begin{proof} 
The proof is immediate by using the relation: 
${\rm vec}({\bf A}{\bf B} {\bf C}) = \left( {\bf C}^T \otimes {\bf A} \right) 
{\rm vec}({\bf B}).$ 
\end{proof} 

It is possible to obtain interesting, yet realistic statistical models 
that allow tractable performance analysis if we make the following 
simplifying assumptions. The following notation is used: The 
vectors ${{\bf g}}_i$  (and ${{\bf h}}_j$) denote the $i$-th (and the 
$j$-th) column of $\bH^H$ (and $\bH$), respectively, i.e.\ $\bH = 
[{{\bf h}}_1 \ldots {{\bf h}}_{N_t}] = [{{\bf g}}_1 \ldots 
{{\bf g}}_{N_r}]^H$. 

\subsection{Canonical Model 1 ($\CMone$)}  We denote by $\CMone$ a channel that 
follows the following two assumptions. 

\noindent {\em Assumption 1:}\; The covariance matrices of all rows of $\bH$ 
have the columns of $\bU_t$ as a set of common eigenvectors. That is, 
$E \left[  {{\bf g}}_i {{\bf g}}_i^H \right] = \bU_t \bLambda_t^{ii} \bU_t^H$ 
for some positive semi-definite diagonal matrix $\bLambda_t^{ii}$. 

\noindent {\em Assumption 2:}\; The cross-covariance matrices of 
the rows of $\bH$ also have a set of common eigenvectors, given by columns
of $\bU_t$, i.e.\ $E \left[  {{\bf g}}_i {{\bf g}}_j^H \right] = 
\bU_t \bLambda_{t}^{ij} \bU_t^H$ for all $i, j, \hspp i \neq j$ and some 
diagonal $\bLambda_{t}^{ij}$. In general, $\bLambda_{t}^{ij}$ need not be 
positive semi-definite because $E \left[  {{\bf g}}_i {{\bf g}}_j^H \right]$ 
is not Hermitian. 

\noindent Then, $\bH$ can be written as $\bH = \bH_t \bU_t^H$ with the 
following properties: 
\begin{itemize}
\item The covariance matrix of each row of $\bH_t$ (denoted by $\{ {\bf g}_{ti} \}$), 
given by $E \left[ {{\bf g}}_{ti}  {{\bf g}}_{ti}^H \right]$, is diagonal. This 
follows directly from Assumption 1 and the fact that 
${{\bf g}}_{ti} = \bU_t^H {{\bf g}}_{i}$.

\item The columns of $\bH_t$ (denoted by $\{ {\bf h}_{ti} \}$) are uncorrelated 
with each other, i.e.\ 
$E \left[ \bhu_{ti} \bhu_{tj}^H \right] = {\bf 0}$ for all $i,j, \hspp i \neq j$. 
However, the columns may have arbitrary covariances. 
This is because the $(m,n)$-th entry of the cross-covariance between the $i$-th 
and the $j$-th columns of $\bH_t$ is given by $E  \left[  {\bf g}_{tm}^{*}[i] 
{\bf g}_{tn}[j] \right]$ which can be seen from Assumptions 1 and 2 to be 
$\delta(i - j) \bLambda_t^{nm}[j]$ with $\delta(\cdot)$ denoting the 
Kronecker delta. 
\end{itemize} 
We summarize these conclusions in the form of Lemma~\ref{lemmaCM1}. 
\begin{lemma} [$\CMone$] 
Under Assumptions 1 and 2, any channel can be written as $\bH = \bH_t \bU_t^H$ 
where the columns of $\bH_t$ are uncorrelated with each other. \label{lemmaCM1}
\endproof
\end{lemma}

\subsection{Canonical Model 2 ($\CMtwo$)}
We denote by $\CMtwo$ a channel which follows the following two assumptions. 

\noindent {\em Assumption 3:}\; The covariance matrix of the columns of $\bH$, 
given by $E \left[ \bhu_{i} \bhu_{i}^H \right]$, have a set of common 
eigenvectors, independent of $i$. We assume that these eigenvectors are 
columns of $\bU_r$. 

\noindent {\em Assumption 4:}\; The cross-covariance matrices of columns of 
$\bH$ also have a common set of eigenvectors, given by columns of $\bU_r$, i.e.\ 
$E \left[ \bhu_{i}  \bhu_{j}^H \right] = \bU_r \bLambda_{r}^{ij} \bU_r^H$ 
for all $i, j, \hspp i \neq j$ and some diagonal $\bLambda_{r}^{ij}$. 

\begin{lemma} [$\CMtwo$] 
Under Assumptions 3 and 4, any channel can be written as $\bH = \bU_r \bH_r$ 
where the rows of $\bH_r$ are uncorrelated with each other. 
\endproof \label{lemmaCM2}
\end{lemma} 
It is clear that $\CMtwo$ is the dual of $\CMone$. In $\CMone$, the transmitter 
sees parallel 
channels in the eigen-domain while in $\CMtwo$, the receiver sees parallel 
channels in the eigen-domain.

\subsection{Canonical Model 3 ($\CM$)} We denote a channel which follows 
Assumptions~$1$-$4$ as $\CM$. This model has been developed independently 
in~\cite{Bonek,tulino_ind,Kotechacaptech}. 
\begin{lemma} [$\CM$] 
Under Assumptions~$1$-$4$, any channel can be written as $\bH = \bU_r \bH_c \bU_t^H$ 
where the entries of $\bH_c$ are uncorrelated, but not necessarily identically 
distributed.
\label{lemmaCM3}
\end{lemma}
\begin{proof}
See Appendix~\ref{app0}. 
\end{proof} 
Exploiting the fact that the entries of $\bH_c$ are uncorrelated under $\CM$, 
the channel covariance matrix can be written as 
\beq 
\label{ref1} 
\bR = (\bU_t^T \otimes \bU_r^H)^H \hsppp \bR_c \hsppp (\bU_t^T \otimes \bU_r^H) 
\eeq 
where $\bR_c$ is diagonal. First, note that the right-hand side of (\ref{ref1}) 
is an eigen-decomposition of $\bR$. The matrices $\bU_t$ and $\bU_r$, which are 
a set of eigenvector matrices of $\bQi_t = E \left[ \bH^H \bH \right]$ and 
$\bQi_r  = E \left[ \bH \bH^H \right]$, can be interpreted as transmit and 
receive eigen-matrices, respectively. Clearly, $\CM$ is a special case of $\CMone$ 
(and $\CMtwo$) where the covariance matrices of the columns (rows) of $\bH_t$ 
($\bH_r$) have the same eigen-matrix $\bU_t$ ($\bU_r$). In fact, $\CM$ is an 
intersection of $\CMone$ and $\CMtwo$. Our primary focus in the rest of the 
paper is on $\CM$, which we will label as the {\em canonical model.} We 
also define the {\em spatial power} matrix $\bP_c$ by the relationship 
\begin{eqnarray} 
\bP_c[i,j] \triangleq E \left[ | \bH_c [i,j]|^2 \right] 
\end{eqnarray} 
and note that the diagonal entries of $\bR_c$ correspond to $\{ \bP_c[i,j] \}$. 
Henceforth, we will use this alternate characterization of $\bR_c$. 
We now identify some of the key properties of $\CM$.

\section{Properties of The Canonical Model} 
\label{cm3_features}
\subsection{Relation to Other Channel Models}
\label{comp} 
We show how two well-known channel models, the Kronecker model and the 
virtual representation framework, can be regarded as special cases of the 
canonical model. 
\subsubsection{Kronecker Model} 
\label{trchannel} 
The Kronecker model has been used in~\cite{Chuah,Shiu} and also in many 
recent works under the assumption that the transmitter and the receiver 
are surrounded by local scatterers. This model is verified by measurement 
campaigns for certain environments 
in~\cite{Kai,mcnamara,kron_int1,wallace,mimo_manhattan}. 
It assumes separable statistics at the transmitter and the receiver and is 
given by 
\beq
\label{channeltr} 
\bH = \Sigmabf_r^{1/2} \hsp \bH_{\iid} \hsp \Sigmabf_t^{1/2}, 
\;\;\;\;\; \;\;\;\; \bR_k \triangleq E \left[ {\rm vec}({\bf H}) 
{\rm vec}({\bf H})^H  \right] = \Sigmabf_t \otimes \Sigmabf_r
\eeq 
where the entries of $\bH_{\iid}$ are i.i.d.\ $\compnorm(0,1)$, and 
$\Sigmabf_t$ and $\Sigmabf_r$ are the transmit and the receive covariance 
matrices, respectively. 

Recall from the discussion in Sec.~\ref{MIMO} that 
\begin{eqnarray} 
\bQi_{t} = E \left[ \bH^H \bH \right] = \bU_t \bLambda_t \bU_t^H,
& &
\bQi_{r} = E \left[ \bH \bH^H \right] = \bU_r \bLambda_r \bU_r^H, 
\\ 
\bR = E\left[ {\rm vec}(\bH) {\rm vec} (\bH)^H  \right], & {\rm and} 
& {\rm Tr}(\bR) = {\rm Tr}(\bQi_t) = {\rm Tr}(\bQi_r). 
\end{eqnarray}
From~(\ref{channeltr}), we then have the following relations for 
the Kronecker model: 
\begin{align} 
\label{kronr} 
\bQi_{t}  = \Sigmabf_t^{1/2} \hspp E \left[ \bH_{\iid}^H \hspp
\Sigmabf_r \hspp \bH_{\iid} \right] \hspp \Sigmabf_t^{1/2} = 
\Sigmabf_t \hspp {\rm Tr}(\Sigmabf_r), \\ \label{kront} 
\bQi_{r}  = \Sigmabf_r^{1/2} \hspp E \left[ \bH_{\iid} \hspp 
\Sigmabf_t \hspp \bH_{\iid}^H \right] \hspp \Sigmabf_r^{1/2} = 
\Sigmabf_r \hspp {\rm Tr}(\Sigmabf_t). 
\end{align} 
The fact that $\bQi_{\bullet}$ and $\Sigmabf_{\bullet}$ are scaled 
versions of each other implies that eigen-decompositions of $\Sigmabf_t$ 
and $\Sigmabf_r$ are given by $\bU_t \bLambda_{t,k} \bU_t^H$ and 
$\bU_r \bLambda_{r,k} \bU_r^H$ where $\bLambda_{t,k} = 
\bLambda_t/{\rm Tr}(\Sigmabf_r)$ and $\bLambda_{r,k} = 
\bLambda_r /{\rm Tr}(\Sigmabf_t)$, respectively. Further, 
\begin{eqnarray} 
{\rm Tr}(\bQi_t) = {\rm Tr}(\bQi_r) = {\rm Tr}(\Sigmabf_t) 
{\rm Tr}(\Sigmabf_r) = {\rm Tr}(\bR). 
\end{eqnarray} 
Thus, we can write the channel in (\ref{channeltr}) in the $\CM$ 
form as 
\beq
\label{equivtr}
\tau \hsppp \bH 
= \bU_r \hspp \bLambda_r^{1/2} \hspp \bU_r^H 
\hspp \bH_{\iid} \hspp \bU_t \hspp \bLambda_t^{1/2} \hspp \bU_t^H  
= \bU_r \hspp \bH_c \hspp \bU_t^H 
\eeq
where $\tau = \left[  {\rm Tr}(\bR) \right]^{1/2}$. The entries of 
${\bH}_c$ are uncorrelated with covariance matrix ${\bf R}_c$ where 
\begin{eqnarray} 
\bH_c & = & \bLambda_r^{1/2} \hspp \bU_r^H \hspp \bH_{\iid} \hspp 
\bU_t \hspp \bLambda_t^{1/2} \stackrel{\mathit {(a)}}{\sim} 
\bLambda_r^{1/2} \hspp \bH_{\iid} \hspp \bLambda_t^{1/2}
\label{eqn_mm1} \\ 
\bR_c & = & 
E \left[ \bh_c \bh^H_c \right] = \bLambda_t \otimes \bLambda_r, 
\hsp \bh_c \triangleq {\rm vec}(\bH_c). 
\end{eqnarray} 
The equality in~(a) of~(\ref{eqn_mm1}) arises from the invariance 
of the distribution of $\bH_{\iid}$ under left and right unitary 
multiplications~\cite{Telatar}.

The spatial power matrix for the Kronecker model is denoted by $\bP_k$. 
The Kronecker structure of $\bR_c$ implies that the $i$-th column vector 
of $\bP_k$ (denoted by $\bP_{k,i}$) is 
\beq
\label{separable}
\bP_{k,i} = \bLambda_t [i] \cdot \Big[ \;\bLambda_r [1], 
\bLambda_r [2], \ldots , \bLambda_r [N_r] \; \Big]^T. 
\eeq
Note that this is a direct consequence of assuming separable statistics 
for $\bH$ and does not hold in general. For the general case, separation 
of the transmit and receive domains can be artificially induced by using 
the {\em marginal} sum statistics (see Prop.~\ref{prop_modelfit}). 
This fact highlights the limitations of the Kronecker model. The canonical 
framework results in a richer class of channels since it does not assume 
separability and an arbitrarily diagonal $\bR_c$ is needed to model a 
general channel. 

\subsubsection{ULAs and the Virtual Representation}
\label{ULA} 
In~\cite{Sayeeddecon}, a virtual representation framework is proposed 
for systems with ULAs at both the transmitter and the receiver. In this 
case, $\bH$ can be written as $\bA_r \bH_v \bA_t^H$ where $\bA_t$ and 
$\bA_r$ are discrete Fourier transform (DFT) matrices. It is argued 
in \cite{Sayeeddecon} that the entries of $\bH_v$ are approximately 
uncorrelated for finite number of antennas and the approximation becomes 
increasingly accurate as antenna dimensions increase. Thus, $\bA_t$ and 
$\bA_r$ serve as eigen-matrices in the virtual representation framework: 
\begin{eqnarray}
E \left[ \bH^H \bH \right] & = & \bA_t \bLambda_{t,v} \bA_t^H, \hsp \hsp 
\bLambda_{t,v} = E \left[ \bH_v^H \bH_v  \right], 
\\ E \left[ \bH \bH^H \right] & = & \bA_r \bLambda_{r,v} \bA_r^H, \hsp \hsp 
\bLambda_{r,v} = E \left[ \bH_v \bH_v^H  \right]. 
\end{eqnarray}

Furthermore, Assumptions~$1$-$4$ made in the context of $\CM$ are satisfied by 
virtual representation. An important point to note is that while the transmit 
and the receive basis in $\CM$ are a function of the channel statistics and the 
entries of the canonical decomposition are {\em exactly} uncorrelated, the 
eigen-matrices $\bA_t$ and $\bA_r$ are fixed DFT matrices and entries of 
$\bH_v$ are {\em approximately} uncorrelated. {\em Thus, in addition to the fact 
that the virtual representation for ULAs provides an intuitive physical 
interpretation where the eigenvectors $\bA_t$ and $\bA_r$ are beams in fixed 
virtual directions, it also makes transmit signal design easier since the 
transmit and the receive bases are fixed and do not change with the channel 
statistics.} 

\subsection{Transmit-Receive Eigen-spaces and Their Interaction}
\label{transmit_rec}
The decomposition of $\CM$ provides an equivalent representation 
in the eigen domain: 
\begin{eqnarray} 
\by_c = \bH_c \bx_c + \bn_c 
\end{eqnarray} 
where 
\begin{eqnarray} 
\by_c \triangleq \bU_r^H \by, \hspp 
\bx_c \triangleq \bU_t^H \bx, \hsp  {\rm and} \hsp 
\bn_c \triangleq \bU_r^H \bn. 
\end{eqnarray} 
Thus, a linear transformation at the transmitter and the receiver 
results in $\bH_c$ with independent entries. We note the following points. 
\begin{itemize} 
\item {\em Joint statistics} - $\bH_c$ captures the {\em joint} transmitter-receiver 
statistics given by $\bP_{c}$ which are in general non-separable, in comparison 
with the separable statistics of $\bP_{k}$. 

\item {\em Degrees of freedom} - Define the DoF available in the channel as the 
entries of $\bH_c$ having non-zero\footnote{In practice, it is reasonable to 
define $\DoF$ as the number of entries in ${\bf P}_c$ that are larger than 
an {\em a priori}-determined threshold. The term ``rank'' in~(\ref{rank_eqn}) 
should then be replaced with an appropriate definition of ``effective rank.''} 
variance. Thus, 
\begin{eqnarray}
\label{rank_eqn}
\DoF = \rank(\bR_c) = \rank(\bR) \leq N_t N_r. 
\end{eqnarray}
The i.i.d.\ channel has $\DoF = N_t N_r$ and all the DoF have equal power. 
In correlated channels, however, the DoF is smaller and these DoF do 
not have equal power. 

\item {\em Parallel channels} - The parallel channels in the i.i.d.\ 
case have identical statistics and number $\min(N_t, N_r)$. In correlated 
channels, the non-zero columns of $\bP_{c}$ expose the number of available 
parallel channels which is less than $\min(N_t, N_r)$, in general. 
\end{itemize}
The last two observations signify the key differences in correlated versus 
i.i.d.\ channel modeling. Since the DoF and parallel channels are unequal, 
they should be excited appropriately for optimal transmission. {\em While 
the canonical model does not provide the same physical insight as the 
virtual representation (e.g., path partitioning), the mathematical 
similarities between the two models can be exploited.} This is witnessed 
by many recent works that explore the impact of independent entries in the 
case of virtual representation. See e.g.,~\cite{Kotechachannelest} for 
channel estimation;~\cite{spl_issue,Veeravallicap,tulino_ind} and 
Sec.~\ref{capacity} of this paper for capacity 
analysis;~\cite{vasanth_limbf,vasanth_limprecode,vas_matched} for 
limited feedback 
system design;~\cite{noncoh_mahesh,noncoh_mahesh2} for non-coherent signal 
design;~\cite{Kotechaglobecomm2003,vasanth_clin_vvv} for space-time code 
design etc.

\section{Statistical Models for Measured Channels} 
\label{sec_prac}
In this work, we adopt the standard channel power normalization used in the 
MIMO literature: $\chanpow = N_t N_r$, where $\chanpow$ is defined as 
\begin{eqnarray} 
\chanpow \triangleq E[ {\rm Tr} (\bH \bH^H)] = 
E[ {\rm Tr} ( \bH_c \bH_c^H ) ] = {\rm Tr}( \bR ) = 
{\rm Tr}(\bR_c) = \sum_{ij} \bP_c[i,j]. 
\end{eqnarray} 

\subsection{Fitting Measured Channels with a Kronecker Model} 
\label{model_fit}
Even though some initial 
studies~\cite{Kai,mcnamara,kron_int1,wallace,mimo_manhattan} indicate that 
the Kronecker model is a good fit for $2 \times 2$ scenarios, further 
studies~\cite{Bonek,zhou,costa_haykin,new_ozcelik,wyne,abhayapala,bonek_valid,survey} 
show that a non-separable modeling framework is more accurate. A 
non-separable framework in a Rayleigh fading setting is characterized 
by $N_t N_r$ statistical parameters, namely $\{ {\bf P}_c[i,j] \}$. 
Initial difficulties on the 
tractability of the performance analysis of MIMO channels with such a 
general statistical description 
has led to the popularity of fitting the measured channel with a model 
characterized by fewer parameters. 
The following 
proposition illustrates how a physical channel generated assuming 
non-separable statistics can be fitted with a Kronecker model. 
\begin{prop} 
\label{prop_modelfit}
Consider a channel under $\CM$: $\bH = \bU_r \bH_c \bU_t^H$ with 
$\bP_c[i,j] = E \left[ | \bH_c [i,j]|^2 \right]$. A Kronecker fit for $\bH$ is 
of the form $\bU_r \bH_k \bU_t^H$ where $\bH_k[i,j] \sim 
\compnorm(0, \bP_k[i,j] )$ with 
{\vspace{0.05in}} 
\begin{eqnarray} 
\label{kron_var}
\bP_{k}[i,j] = \frac{\sum_l \bP_{c}[i,l] \cdot \sum_k \bP_{c}[k,j]  } 
{ \sum_{kl} \bP_{c}[k,l] }. 
\end{eqnarray}
Furthermore, the mapping in~(\ref{kron_var}) {\em always} increases 
the DoF in the Kronecker fit for a scattering environment described 
by $\CM$. 
\end{prop} 
{\vspace{0.05in}}
\begin{proof} 
Given a channel $\bH$ that follows $\CM$, we attempt to fit a channel $\widetilde{\bH}$ 
that follows the Kronecker model to it. From Sec.~\ref{comp}, the general 
form of $\widetilde{\bH}$ is $\bU_{rk} \hspp {\bf \Lambda}_{rk}^{1/2} \hspp 
\bH_{\iid} \hspp {\bf \Lambda}_{tk}^{1/2} \hspp \bU_{tk}^H$ for some appropriate 
choice of $\bU_{tk}$, $\bU_{rk}$, ${\bf \Lambda}_{tk}$ and ${\bf \Lambda}_{rk}$. 
By comparing the transmit and the receive covariance matrices with the two 
expansions, it can be checked that $\bU_{tk} = \bU_t$, $\bU_{rk} = \bU_r$ and 
$\bP_k[i,j] \triangleq  \bLambda_{rk}[i] \hspp \bLambda_{tk}[j]$ has to satisfy 
the relationship in~(\ref{kron_var}). For the second part, note 
that 
\begin{eqnarray}
{\bf P}_k[i,j] \geq \frac{ \left( {\bf P}_c[i,j] \right)^2 } 
{  \sum_{kl} {\bf P}_c[k,l]  } 
\end{eqnarray}
and hence, ${\bf P}_k[i,j]$ is non-zero if ${\bf P}_c[i,j]$ is. 
Thus, the DoF in the Kronecker fit is always larger than the actual 
DoF with the canonical model. 
\end{proof}

As an extreme artificial example of the above trend, consider a 
$4 \times 4$ system where the $\CM$ channel has $\DoF = 4$ and 
spatial power matrix $\bP_{c}$ as in~(\ref{smoothing}) below. It 
maps to a Kronecker model with $\DoF = 16$ and spatial power matrix 
$\bP_{k}$ as below: 
\begin{align}
\label{smoothing}
\bP_{c} = 4 * \left[
             \begin{array}{cccc}
               1 & 0 & 0 & 0 \\
               0 & 1 & 0 & 0 \\
               0 & 0 & 1 & 0 \\
               0 & 0 & 0 & 1 \\
             \end{array}
           \right]; \;\;\;
\bP_{k} = 
\left[
             \begin{array}{cccc}
               1 & 1 & 1 & 1 \\
               1 & 1 & 1 & 1 \\
               1 & 1 & 1 & 1 \\
               1 & 1 & 1 & 1 \\
             \end{array}
           \right].
\end{align} 
{\emph{In general, the Kronecker model spreads the degrees of freedom
across the resulting $\bP_{k}$ and thereby `flattens' it since its
statistics are based only on column and row sum statistics of the
actual spatial power matrix.}} Note that while the transformation 
from ${\bf R}_c$ to ${\bf R}_k$ could lead to a change in rank, the 
transformation\footnote{As long as no row or column of ${\bf P}_c$ (and 
${\bf P}_k$ following Prop.~\ref{prop_modelfit}) are completely zero, 
the event where ${\bf H}_c$ (similarly for ${\bf H}_k$) is singular is a 
zero probability event~\cite{girko}.} 
from ${\bf H}_c$ to ${\bf H}_k$ does not.

\subsection{Modeling Sparsity Mathematically} 
\label{sec_sparse} 
Another property suggested by fundamental electromagnetic 
studies~\cite{bucci,migliore,ada,massimo,xu_janaswamy,hanlen_sparse,rod_kennedy_sparse,goodman} 
as well as recent measurement 
campaigns~\cite[Figs.\ 9 and 11]{bonek_valid},~\cite{wood_hodgkiss,costa_haykin,zhou} 
is that only a small subset of the $N_t N_r$ statistical parameters in 
$\CM$ are dominant enough to be leveraged towards reliable communications 
over practical $\snr$ ranges. That is, measured wireless channels are 
{\em sparse}. 

In this work, we compare the trends of the canonical and the Kronecker 
models across a large family of correlated$/$sparse channels. We 
provide two simple mathematical frameworks to generate large 
families of channel correlation information under $\CM$ and hence, 
from Prop.~\ref{prop_modelfit}, under the Kronecker model. For this, 
we write $\cvarij$ as 
\begin{eqnarray}
\label{eqn_ref2}
\cvarij = N_t N_r \cdot 
\frac{\punnormij } { \sum_{ij} \punnormij} 
\end{eqnarray}
where $\{ \punnormij \}$ is a family of $N_t N_r$ random variables 
supported on $[0,1]$ that correspond to unnormalized\footnote{That is, 
$p_{i,j}$ have to be normalized, as in~(\ref{eqn_ref2}), to ensure that 
$\rho_c = N_t N_r$.} variances. 

In sparse framework {\sf I}, we set\footnote{The i.i.d.\ assumption on 
$\{ p_{i,j} \}$ is made to simplify further analysis.} $\{ p_{i,j} \}$ 
to be i.i.d.\ with common mean and variance, $\mu$ and $\sigma^2$, 
respectively. A typical rich environment (intuitively, a `near-i.i.d.' 
environment) is obtained by setting $\sigma^2 \approx 0$ with an i.i.d.\ 
channel corresponding to the extreme case of $\sigma^2 = 0$. As $\sigma^2$ 
increases, subject to the condition that $\sigma^2 \leq 1 - \mu^2$ (since 
$p_{i,j}$ are supported on $[0,1]$), $\{ \punnormij \}$ get `well-spread 
out' around $\mu$. That is, there exists a large variability in the 
values of $\{ \cvarij \}$, which intuitively reflects a 
correlated$/$sparse setting. 

Despite a precise recipe for modeling in framework {\sf I}, 
it could be difficult to systematically generate extremely sparse channels 
(where the fraction of dominant entries vanishes). In such settings, 
we propose sparse framework {\sf II} in which we set $p_{i,j}$ as 
\begin{eqnarray}
\label{newe2}
p_{i,j} = q_{i,j} s_{i,j} 
\end{eqnarray}
where $q_{i,j}$ is generated as described above (in framework {\sf I}) and 
$s_{i,j}$ is an i.i.d.\ family of binary ($0$ or $1$)- valued random 
variables with 
\begin{eqnarray}
{\rm Pr}(s_{i,j} = 1) = p = 1 - {\rm Pr}(s_{i,j} = 0). 
\end{eqnarray} 
Sparse channels can be generated systematically by adjusting the value 
of $p$ appropriately. As $p$ increases, the channel generated 
via~(\ref{newe2}) becomes {\em more richer} with the two frameworks 
coinciding for $p = 1$. 

Note that frameworks {\sf I} and {\sf II} 
provide simple mathematical abstractions to model sparsity and their 
applicability in practice needs to be substantiated with further 
measurement studies. Nevertheless, as we will see, these simple models 
provide engineering intuition on the trends of capacity behavior. 

\section{Capacity of Correlated MIMO Channels} 
\label{coh_cap_norm} 
Towards this goal, we now briefly summarize some of the recent works 
on MIMO capacity. Prior to this summary, we state the channel 
state information (CSI) assumptions of this work. 

\subsection{Channel State Information}
We assume a coherent receiver architecture. That is, the receiver has 
perfect CSI. This is possible in practice 
by estimating the channel at the receiver using 
training symbols over a dedicated training period that lasts a significant 
portion of the channel coherence duration. We further 
assume that the statistics of the channel do not change over a 
reasonably long duration so that they can be acquired perfectly at 
the transmitter.

\subsection{Ergodic Capacity}
In this setting, the ergodic (or average) capacity at a transmit $\snr$ 
of $\rho$ is given by~\cite{Foschini} 
\beq 
\label{cap} 
C_{\erg}(\rho) = \sup_{{\bf Q} \; : \; {\bf Q} \hsppp 
\geq \hsppp {\bf 0}, \; \; 
{\rm Tr}({\bf Q}) \hsppp \leq \hsppp \rho} 
\;\; E_{\bH} \left[ 
\log_2  \det \left( \bI +  {\bf H} {\bf Q} {\bf H}^H  \right)
\right] 
\eeq 
where the optimization is over the set of trace-constrained, positive 
semi-definite matrices. While uniform-power (or full rank) signaling is 
optimal when no CSI is available at the transmitter, it is shown 
in~\cite{Visotsky,Tulino,liang} that the optimal ${\bf Q}$ 
to solve (\ref{cap}) has an eigen-decomposition 
\begin{eqnarray} 
{\bf Q}_{\opt} = \bU_t \bLambda_{\opt} \bU_t^H 
\end{eqnarray}
where $\bU_t$ is an eigen-matrix of $\bQi_t = E \left[\bH^H \bH \right]$ 
and $\bLambda_{\opt}$ is a 
positive semi-definite, diagonal matrix obtained via a numerical search. 
Closed-form solutions for $\bLambda_{\opt}$ are not known; however, an iterative 
algorithm has been proposed in~\cite{Tulino}. 

For any correlated channel, this algorithm converges to beamforming (or 
$\rank$-$1$ signaling) in the asymptotically low-$\snr$ regime and 
uniform-power signaling\footnote{Without any loss in generality, we 
assume that no column of $\bP_c$ has all zero entries.} in the 
asymptotically high-$\snr$ regime. Thus, the low-$\snr$ and the 
high-$\snr$ ergodic capacities 
(denoted by $C_{\low}(\rho)$ and $C_{\high}(\rho)$, respectively) are given by 
\begin{eqnarray} 
\label{defn_lowsnr}
C_{\low}(\rho) \triangleq 
C_{\erg}(\rho)\Big|_{ {\bf Q}_{\sf opt} \hspp {\rm as} \hspp 
\rho \hspp \rightarrow \hspp 0 } & = & 
E \left[ \log _2 \bigg( 1 + \rho \sum_{i} \big| \bH _c[i, j_{\max}] 
\big|^2 \bigg) \right] \label{lowsnr} \\ 
C_{\high}(\rho) \triangleq 
C_{\erg}(\rho) \Big|_{ {\bf Q}_{\sf opt} \hspp {\rm as} \hspp 
\rho \hspp \rightarrow \hspp \infty } 
& = & E \left[ \log_2 \det \bigg( \bI_{N_r} + \frac{\rho}{ \rank({\bf P}_c )} 
\hsppp \bH_c \bH_c^H \bigg) \right] 
\label{highsnr}
\end{eqnarray} 
where $j_{\max} = \arg \max _j \sum_{i} \cvarij$ corresponds to the 
dominant transmit eigen-direction. 
In general, at an intermediate $\snr$, the optimal rank of ${\bf Q}$ is 
non-decreasing (as $\rho$ increases) with precise estimates available for the 
transient-$\snr$'s when a particular rank signaling scheme becomes 
optimal~\cite{vasanth_isit07}.

\subsection{Outage Capacity} 
It is also well-understood that the ergodic capacity is an insufficient 
metric to understand the fundamental impact fading has on achievable data 
rates and the notion of outage capacity~\cite{Foschini,wyner}~at an outage 
probability of $q \hspp \%$ is relevant. The outage capacity is the maximum 
rate that is guaranteed for at least $(100-q) \%$ of the channel 
realizations and is defined as 
\begin{eqnarray}
\label{outage_cap1}
C_{\out,\hspp q}(\rho) \triangleq \sup_{R \hspace{0.02in} \geq \hspace{0.02in} 0} 
\left(  R \right)  \hsp \hsp \hsp  {\mathrm{s.t.}}  \hsp \hsp \hsp 
{\mathrm{Pr}}\left( \log_2 \det\left[ \bI + {\mathbf{H}} {\bQi} {\mathbf{H}}^H   
\right] < R \right) \leq \frac{q}{100}.   
\end{eqnarray} 
Many recent works have shown that Gaussian approximations to 
$C_{\out, \hspp q}(\rho)$ with mean and variance given by $C_{\erg}(\rho)$ 
and $V(\rho)$, the variance of capacity, are accurate in the large-system 
limit; see~\cite{it_rs01} and references therein. Thus, in the 
large-system limit, $C_{\out, \hspp q}(\rho)$ can be efficiently approximated as  
\begin{equation}
\label{kam}
C_{\out, \hspp q}(\rho) = C_{\erg}(\rho) -  x_q \sqrt{V(\rho)} + \littleo(1)
\end{equation} 
where $x_q$ is the unique solution to 
${\text{erfc}}\big(x_q/\sqrt{2} \big) = 2q$ 
with ${\mathrm{erfc}}(\cdot)$ denoting the complementary error function.

\section{Comparative Study of Capacity of Kronecker and Canonical Models} 
\label{capacity} 
An important point to note from~(\ref{kam}) is that the 
outage capacity is determined upon knowledge of $C_{\erg}(\rho)$ and 
$V(\rho)$. The main focus of this section is thus on understanding 
$C_{\erg}(\rho)$ and $V(\rho)$ when a MIMO channel (with non-separable 
statistics) is fitted with a Kronecker model 
as in Prop.~\ref{prop_modelfit}. We denote by 
$C_{\erg, \hspp \can }(\rho)$ 
and $C_{\erg,\hspp \kron }(\rho)$, the means of capacity under the 
two models and by $V_{\can }(\rho)$ and $V_{\kron }(\rho)$, the 
variances of capacity under these models. 

In this section, we provide good estimates for the above quantities 
under certain conditions. 
While an analytical understanding of these quantities for all $\snr$s 
seems difficult, it is possible to obtain engineering intuition by 
studying the mismatches (between the two capacities) at the low- and 
the high-$\snr$ extremes under a large-system assumption. Since the 
convergence to the large-system regime is typically fast (see 
e.g.,~\cite{it_rs01} and references therein which point out that good 
agreement is possible even with $4$ or $8$ antennas) we expect this 
analysis to be useful in making meaningful conclusions in the finite 
antenna regime.

\subsection{Low-$\snr$ Extreme} 
As noted in Sec.~\ref{coh_cap_norm}, beamforming to the statistically 
dominant\footnote{Without loss in generality, let all the $N_r$ entries 
in the dominant column $\{ {\bf P}_c[i,j_{\max}], i = 1, \cdots, N_r \}$ 
be non-zero.} transmit eigen-mode (which is the same irrespective of 
whether beamforming is done based on the statistics of 
$\bH_c$ or $\bH_k$) is optimal from an ergodic capacity perspective in 
the low-$\snr$ regime. However, many works in the literature define the 
low-$\snr$ regime {\em imprecisely} as ``$\rho \rightarrow 0$.'' 
It is useful to define a transient-$\snr$, $\rho_{\low}$, such 
that beamforming is capacity-optimal if $\rho < \rho_{\low}$. 
Some works (see~\cite{Veeravallicap,Goldsmith} and references therein) 
define $\rho_{\low}$ {\em implicitly} in terms of means of certain 
random variables that are related to $\bP_c$, but are nevertheless 
difficult to compute in closed-form. In~\cite{vasanth_isit07}, using 
tools\footnote{Also, see~\cite{it_rs06} which points this out from a 
reconfigurable antennas point-of-view.} from random matrix theory, it 
is shown that 
\begin{eqnarray}
\rho_{\low} \approx \frac{1}{\sum_{i=1}^{N_r} \bP_c[i,j_{\max}]}. 
\end{eqnarray}

\noindent {\bf \em Capacity Computation:} We first develop a general 
low-$\snr$ characterization of MIMO capacity in the canonical case, and 
then leverage this result to the Kronecker case. 
For this, we define\footnote{The difference between 
$\rho_{\low, \hsppp \can}$ and $\widehat{\rho}_{\low, \hsppp \can}$ is 
that while beamforming is {\em exactly} capacity-optimal below 
$\rho_{\low, \hsppp \can}$, it is only near-optimal below 
$\widehat{\rho}_{\low, \hsppp \can}$. Nevertheless, 
note that if $\bP_c[i,j_{\max}] = 1$ for all $i$, 
$\widehat{\rho}_{\low, \hsppp \can}$ reduces to $\frac{2}{3N_r}$ and 
thus the trends of $\widehat{\rho}_{\low,\hsppp \can}$ are similar to 
that of $\rho_{\low, \hsppp \can}$.} 
$\widehat{\rho}_{\low, \hspp \can}$: 
\begin{eqnarray} 
\widehat{ \rho}_{\low, \hspp \can} \triangleq 
\frac{1} {\sum_{i=1}^{N_r} \bP_c[i,j_{\max}] } 
\cdot \frac{1}{\gamma_0}, \hsp \hsp 
\gamma_0    =  1 +  \frac{ \sqrt{N_r \sum_{i=1}^{N_r} (\bP_c[i,j_{\max}])^2  }  } 
{2 \sum_{i = 1}^{N_r} \bP_c[i,j_{\max}] }.  
\label{gamma00}
\end{eqnarray} 
The importance of $\widehat{\rho}_{\low, \hspp \can}$ is that $I(\rho)$, 
the average mutual information with statistical beamforming, is given by 
\begin{eqnarray} 
I(\rho) = \log_2(e)\cdot  \rho \cdot \sum_{i=1}^{N_r} \bP_c[i,j_{\max}] 
\cdot(1 + \littleo(1)), \hsp \hsp \rho < \widehat{\rho}_{\low, \hspp \can}. 
\end{eqnarray} 
It should be noted that $C_{\low}(\rho)$ shows the same trends as $I(\rho)$. 
This 
is the content of the following theorem. 
\begin{theorem} 
\label{mv_lowsnr}
There exist positive constants $c, \ell > 0, m > 1/2$ (all independent of 
$\bP_c, N_t$ and $N_r$) such that 
\begin{eqnarray}
\label{mean_var_eqn} 
\log_2(e) \cdot \delta 
\left(1 - \frac{2^{\ell} \sqrt{\kappa_c} }{N_r^m } - 
\frac{\kappa_c \delta}{\gamma_0}  
\right) 
\leq & C_{\erg, \hspp \can}(\rho) & \leq 
\log_2(e) \cdot \delta \nonumber \\ 
& \hsp {\rm for} \hsp {\rm all} \hsppp & {\hspace{-0.02in}}
\rho = \frac{\delta }{ \sum_{i=1}^{N_r} \bP_c[i,j_{\max}] }, 
\hspp 0 < \delta < c, 
\end{eqnarray}
where $\kappa_{c}$ is defined as 
\begin{eqnarray} 
\label{kappa}
\kappa_{c}  \triangleq   
1 + \frac{ \sum_{i=1}^{N_r} \left (\bP_c[i,j_{\max}] \right)^2  } 
{ \left( \sum_{i=1}^{N_r} \bP_c[i,j_{\max}] \right)^2 } 
\end{eqnarray} 
and $\gamma_0$ is as in (\ref{gamma00}). Alternately, the above statement 
can be recast as 
\begin{eqnarray} 
\label{lowsnr_eqn_closedform}
C_{\erg, \hspp \can}(\rho) = 
\log_2(e) \cdot \rho \cdot \sum_{i} \cvarijmax \left(1 + \littleo(1) \right) 
\end{eqnarray}
with the $\littleo(1)$ factor converging to $0$ as $N_r \rightarrow \infty$ and 
$\delta \rightarrow 0$. 
\end{theorem}
\begin{proof}
See Appendix~\ref{append:app1}. 
\end{proof} 

From~(\ref{kron_var}), it follows that if $\{ {\bf P}_c[i,j_{\max}] \}$ 
are non-zero, so are $\{ {\bf P}_k[i,j_{\max}] \}$. 
It is thus easy to specialize Theorem~\ref{mv_lowsnr} to the 
Kronecker case (associated with $\widehat{\rho}_{\low, \hspp \kron}$) 
and compare the two results. 

\noindent{\bf \em Capacity Comparison:} 
\begin{theorem}
\label{mv_cankron} 
Let the low-$\snr$ regime be defined as $\rho < \widehat{ \rho}_{\low} $ 
where 
\begin{eqnarray} 
\widehat{\rho}_{\low} \triangleq 
\min \left( \widehat{\rho}_{\low, \hspp \can}, 
\widehat{\rho}_{\low, \hspp \kron} \right). 
\end{eqnarray}
In this regime, the following conclusions hold for the dominant terms 
of the capacity quantities. 
\begin{itemize} 
\item 
(a) The dominant terms of the ergodic capacity under the two models 
is the same. In particular, we have 
\begin{eqnarray} 
C_{\erg, \hspp \can }(\rho) = C_{\erg, \hspp \kron }(\rho) 
& = & \log_2(e) \rho \cdot \sum_{i} {\bf P}_{c }[i, j_{\max}]. 
\end{eqnarray} 
\item 
(b) The dominant terms of the variances satisfy 
\begin{eqnarray} 
\frac{ V_{\can  }(\rho)} { \left(\log_2(e) \rho \right)^2 } 
= \sum_{i} \cvarijmaxfour , & & 
\frac{ V_{\kron }(\rho) } {\left( \log_2(e) \rho \right)^2} = 
\sum_{i} \kvarijmaxfour. 
\label{var_1}
\end{eqnarray} 
\item 
(c) Let ${\bf P}_c$ be row-permuted such that 
$\{ \sum_{k=1}^{N_t} {\bf P}_c[i,k] , \hsp i = 1, \cdots , N_r \}$ is 
arranged in decreasing order. Further, if the entries of ${\bf P}_c$ 
satisfy 
\begin{eqnarray}
\label{ass_yixi} 
\frac{ \sum_{k \neq j_{\max} } {\bf P}_c[i,k]  }
{ \sum_{k \neq j_{\max} } {\bf P}_c[i+1, k] } \leq 
\frac{ {\bf P}_c[i, j_{\max}] }{ {\bf P}_c[i+1,j_{\max}] } 
\hsp {\rm for} \hsp {\rm all} \hsp 1 \leq i \leq N_r - 1, 
\end{eqnarray}
then $V_{\can}(\rho) \geq V_{\kron}(\rho)$ as $\rho \rightarrow 0$. 
\end{itemize} 
\end{theorem}
\begin{proof} 
See Appendix~\ref{app_cankron}. 
\end{proof} 

The condition in~(\ref{ass_yixi}) implies that the fraction of power 
captured in the beamforming direction by a receiver decreases in the 
same order as the {\em total} power captured by the receivers. For 
example, in the case of regular channels (see 
Footnote~\ref{footnote_regular}), it is easy to check 
that~(\ref{ass_yixi}) holds trivially. In fact, for regular channels, 
it can be checked that 
\begin{eqnarray}
\frac{ V_{\sf can}(\rho) }{ V_{\sf kron}(\rho) } = 
\frac{N_r \sum_{i=1}^{N_r} 
\left( {\bf P}_c[i, j_{\max} ] \right)^2  }
{ \left( \sum_{i=1}^{N_r} {\bf P}_c[i, j_{\max} ] 
\right)^2  } \geq 1
\end{eqnarray} 
due to the Cauchy-Schwarz inequality. 
It also seems like the condition in~(\ref{ass_yixi}) is necessary to 
ensure that $V_{\can}(\rho) \geq V_{\kron}(\rho)$. For example, it can 
be checked that $V_{\can}(\rho) < V_{\kron}(\rho)$ in the following 
$2 \times 2$ case where 
\begin{eqnarray}
{\bf P}_c = \left[ 
\begin{array}{cc} 
1 & A(1 + \epsilon) \\ 1 & A 
\end{array} \right], \hsp A \leq \frac{2}{2+ \epsilon}, \hsp 
\epsilon > 0 
\end{eqnarray} 
and~(\ref{ass_yixi}) does not hold. Nevertheless, in the large-system 
regime, we have the following conclusions for the probabilistic sparse 
frameworks {\sf I} and {\sf II}, introduced in Sec.~\ref{sec_sparse}.

\begin{prop}
\label{prop_as} 
First, recall that {\sf I} is a special case of {\sf II} with $p = 1$. 
\begin{itemize}
\item (a) The probability with which the condition in~(\ref{ass_yixi}) 
holds converges to $1$ as $\{N_t, N_r  \} \rightarrow \infty$. Hence, 
$V_{\can}(\rho) \geq V_{\kron}(\rho)$ for ``almost 
all''\footnote{\label{fn_tech}Technically, this statement 
has to be read as: ``with probability 1 on the probability space 
corresponding to $\{ p_{i,j} \}$.'' Henceforth, we will not bother 
with this detail.} sparse scattering 
environments generated from either framework. 

\item (b) In particular, if $0 < m \leq q_{i,j} \leq M$ with 
$E[ q_{i,j} ] = \mu$ and ${\rm Var}( q_{i,j} ) = \sigma^2$, we have 
\begin{eqnarray}
\label{lowsnr_var_bds}
1 \leq 
 \frac{V_{\sf can}(\rho)} 
{V_{\sf kron}(\rho) }  \leq \frac{1}{p} \cdot \frac{ (M+m)^2  }{4 M m} .
\end{eqnarray}
More specifically, we have $\frac{ V_{\can}(\rho) }{ V_{\kron}(\rho) } 
\rightarrow \frac{1}{p} \left( 1 + \frac{\sigma^2}{\mu^2} \right).$

{\vspace{0.05in}}
\item (c) Equality in the lower bound of~(\ref{lowsnr_var_bds}) is 
achieved when ${\bf H}$ is i.i.d. If $N_t = N_r = N$ such that 
$\frac{MN}{M+m}$ and $\frac{m N}{M+m}$ are integers, 
equality in the upper bound is approached 
as $N \rightarrow \infty$ by 
\begin{eqnarray}
\label{pc_ubound}
{\bf P}_c^{\sl T} = \frac{N}{M+m} \cdot 
\left[ 
\begin{array}{c} 
{ \begin{array}{c ccc ccc} 
& M+m &
\underbrace{ m \hsp  \hsp \cdots  \hsp \hsp m }_{ \frac{MN}{M+m} - 1 } 
& 
\underbrace{ M \hsp \hsp  \cdots \hsp \hsp M}_{ \frac{mN}{M+m} } 
& &  & \\ 
\end{array} } \\ 
\begin{array}{c ccc ccc}
0 & M & & & 0 & \cdots & 0 \\ 
\vdots & & \ddots & & \vdots & \ddots & \vdots \\ 
0 & & & M & 0 & \cdots & 0 \\ 
0 & 0 & \cdots & 0 & m & & \\ 
\vdots & \vdots & \ddots & \vdots & & \ddots & \\ 
0 & 0 & \cdots & 0 & & & m 
\end{array}
\end{array}
\right]. 
\end{eqnarray}
\end{itemize}
\end{prop}
{\vspace{0.1in}}
\begin{proof}
See Appendix~\ref{app_prop_as}. 
\end{proof}

Note that the channel corresponding to ${\bf P}_c$ in~(\ref{pc_ubound}) 
is such that ${\bf \Sigma}_t$ has at least $N \left( 1 - \frac{m}{M} \right)$ 
dominant eigenvalues whereas the eigenvalues of ${\bf \Sigma}_r$ are all 
equal to $M+m$. It is surprising that channels that are `near-well-conditioned' 
on both the transmitter and the receiver sides (${\bf H}_{\sf iid}$ 
and the channel in~(\ref{pc_ubound})) could either maximize 
or minimize $\frac{V_{\sf can}(\rho)}{V_{\sf kron}(\rho)}$ depending on 
the distribution of non-zero entries in ${\bf P}_c$. 

\noindent {\bf \em Discussion:} The above results show that the ergodic 
capacities remain the same under the canonical and the Kronecker models 
for all channels in the low-$\snr$ regime. Thus, the dominant factors in 
understanding outage capacity (rate vs.\ reliability trade-off) 
in~(\ref{kam}) are 
the variances of capacity. Since $V_{\can }(\rho) \geq V_{\kron }(\rho)$ 
for almost all sparse channels, 
the outage capacity under the Kronecker model is always steeper than 
the outage capacity under the canonical model (except for i.i.d.\ 
${\bf H}_c$ where they are equally steep). Furthermore, the 
differential in steepness increases as the channel becomes more sparse. 

\begin{figure}
\centerline{\psfig{figure=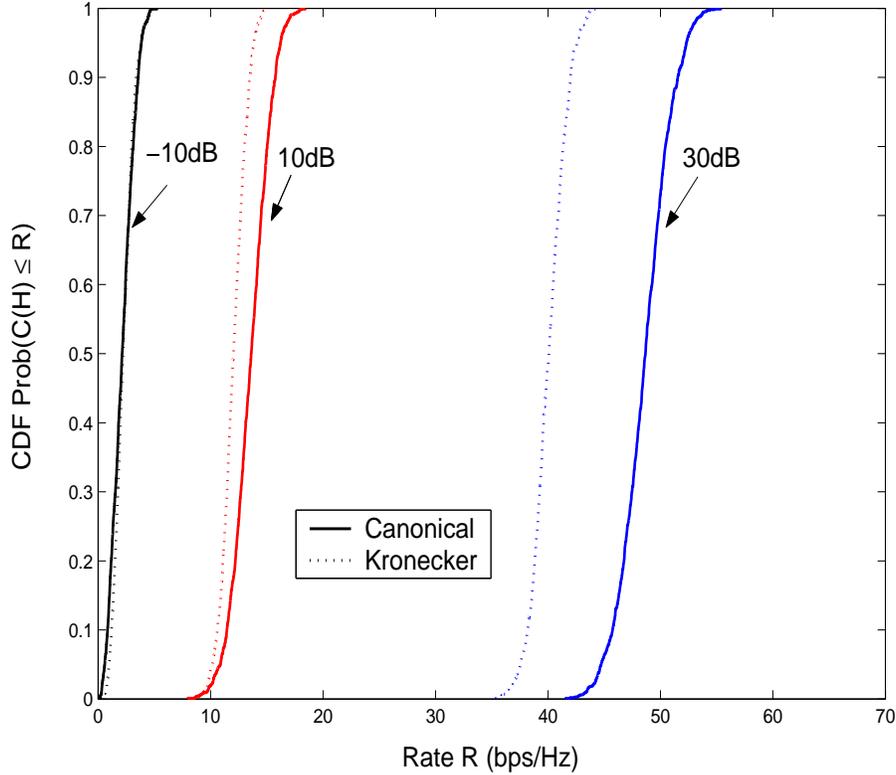,width=4.7in,height=4.1in}}
\caption{Capacity CDFs of a {\em sparse} channel with canonical and 
Kronecker models at $-10, 10$ and $30$ dB $\snr$s.}
\label{fig_cdf_sparse}
\end{figure}

\begin{figure}
\centerline{\psfig{figure=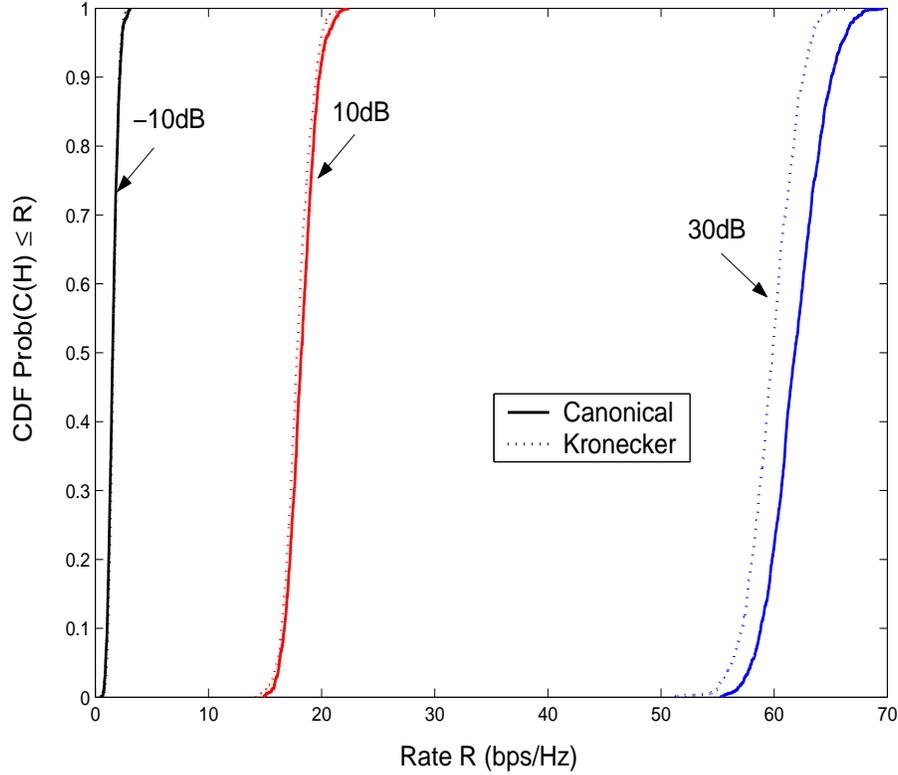,width=4.7in,height=4.1in}}
\caption{Capacity CDFs of a channel that has an intermediate level 
of richness with canonical and Kronecker models at $-10, 10$ and $30$ dB 
$\snr$s.}
\label{fig_cdf_med}
\end{figure}

\begin{figure}[htb!]
\centerline{\psfig{figure=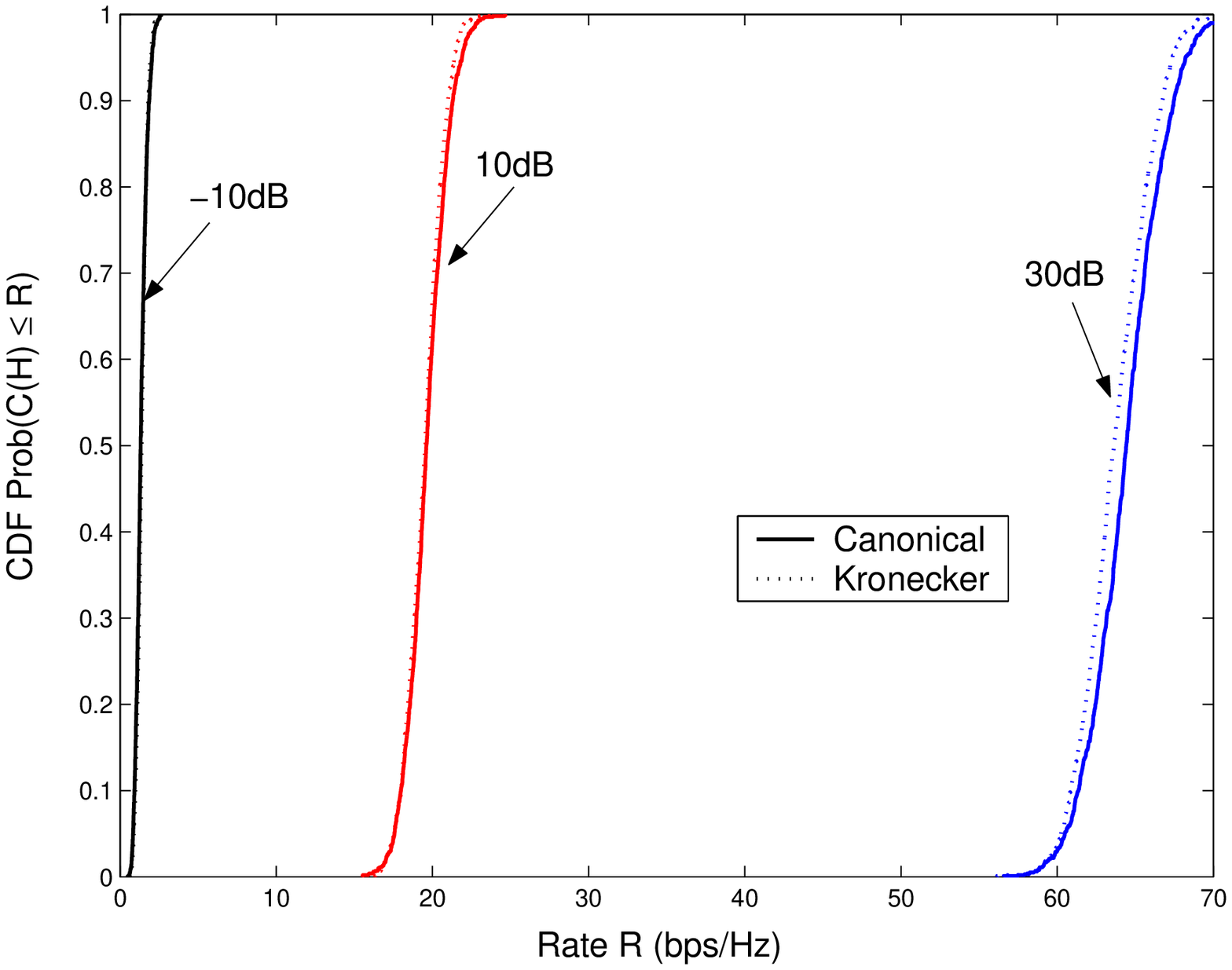,width=4.7in,height=4.1in}}
\caption{Capacity CDFs of a {\em rich} channel with canonical and 
Kronecker models at $-10, 10$ and $30$ dB $\snr$s.}
\label{fig_cdf_rich}
\end{figure}

In other words, at high levels of operational reliability, the Kronecker 
model overestimates capacity while it switches roles and underestimates 
capacity at low levels of reliability. However, the smallness of the 
capacity values generally means that these trends are not prominent when 
we plot outage capacity in the low-$\snr$ regime. For example, 
Figs.~$\ref{fig_cdf_sparse}$-$\ref{fig_cdf_rich}$ plot the cumulative 
distribution function (CDF) of capacity (at $-10, 10$ and $30$ dB $\snr$s) 
for three $8 \times 8$ channels 
generated\footnote{The spatial power matrices for this experiment have 
been generated artificially to mimic certain typical scattering 
environments, and not using the sparse frameworks of 
Sec.~\ref{sec_sparse}.} to portray: i) A typical sparse setting, ii) A 
setting with intermediate level of richness, and iii) A typical rich 
setting. The spatial power matrices are given by 
\begin{align}
\begin{small}
\bP_{c, \hspp {\sf sparse} } = \left[
  \begin{array}{cccccccc}
   43.2693 &   2.5367  &  0.4362  &  0.5569  &  0.1004  &  0.1805  &
   0.1413 & 0.2009 \\
    3.1184 &   1.9519  &  2.4028  &  1.4193  &  0.1007  &  0.1635  &
    0.2420 & 0.1864 \\
    0.4997 &   0.6135  &  0.5352  &  0.5998  &  0.0502  &  0.4960  &
    0.1892 & 0.1532 \\
    0.0574 &   0.0671  &  0.3255  &  0.2233  &  0.0653  &  0.1082  &
    0.1258 & 0.0695 \\
    0.0702 &   0.1138  &  0.1534  &  0.0767  &  0.0327  &  0.0482  &
    0.1087 & 0.0426 \\
    0.0192 &   0.0564  &  0.0540  &  0.1060  &  0.0908  &  0.0608  &
    0.0512 & 0.0426 \\
    0.0147 &   0.0174  &  0.0090  &  0.0132  &  0.0257  &  0.0266  &
    0.0287 & 0.0181 \\
    0.0037 &   0.0032  &  0.0026  &  0.0034  &  0.0029  &  0.0050  &
    0.0038 & 0.0036 \\
  \end{array}
\right]
\label{reallowcan}
\end{small}
\end{align}
while for rich scattering, it is 
\begin{align}
\begin{small}
\bP_{c, \hspp {\sf rich}} = \left[ 
               \begin{array}{cccccccc}
 8.6986  &  3.7188  &  0.7361 &   1.1644 &   2.1912  &  1.4959 &   1.4674 & 1.6262 \\     2.5481  &  3.5585 &   0.5683  &  0.7660  &  1.4141  &  1.3985 &   1.3766 &  1.2021 \\
 0.9910  &  2.1553 &   0.2217  &  0.3105 &   3.0150  &  0.6737 &   0.6415 &  1.7003 \\ 
 0.0793  &  0.1347 &   0.0685  &  0.0750 &   0.1279  &  0.1439 &   0.1218 &  0.0778 \\ 
 0.1651  &  0.1946 &   0.0952  &  0.2151 &   0.3063  &  0.2179 &   0.2742 &  0.2119 \\
 1.2612  &  0.9009 &   0.3837  &  0.6108 &   1.3185  &  0.7591 &   0.9227 &  1.7308 \\
 0.3696  &  1.0783 &   0.4618  &  0.7504 &   0.5129  &  0.5620 &   1.4507 &  0.7081 \\
 0.7332  &  0.2221 &   0.1895  &  0.3626 &   0.6765  &  0.4681 &   0.6841 & 0.7337 
\end{array}
             \right], 
\end{small}
\label{realhighcan}
\end{align} 
and $\bP_{c, \hspp {\sf intermediate}} =  
\frac{\bP_{c,\hspp \sparse} + \bP_{c, \hspp \rich} }{2}$.

Note that $\sum_{ij} {\bf P}_{c, \hspp \bullet} = 64$ for all the three 
channels and the ratio of the largest to the smallest transmit eigenvalue 
decreases from $100.4$ to $9.31$ and $5.45$ as the channel becomes 
progressively richer. The ratio of the largest to the smallest receive 
eigenvalue decreases from $1682$ to $36.6$ and $25.5$ as the channel becomes 
richer. The channel realizations are generated as 
\begin{eqnarray} 
{\bf H}_{\bullet} = {\bf H}_{\iid} \odot (\bP_{c, \hspp \bullet} )^{1/2} 
\end{eqnarray}
where $\bH_{\iid}$ is an i.i.d.\ channel and 
$(\bP_{c, \hspp \bullet} )^{1/2}$ is the element-wise square-root of the 
spatial power matrix. 

\noindent {\bf \em Spectral Efficiency:} 
Another characterization of low-$\snr$ performance is in the context 
of spectral efficiency~\cite{VerduIT2002} (equivalently, 
$C_{\erg,\hspp \bullet}(\rho)$ vs.\ $\rho$ behavior). 
We now present the connections between the canonical and the Kronecker 
models to the two key figures-of-merit in low-$\snr$ communications: 
i) Minimum energy per bit necessary for reliable communication, 
$\sebnomin$, and ii) Wideband slope, $S_0$. For a multi-antenna channel, 
these two metrics are given by~\cite{VerduIT2002}
\begin{eqnarray}
\sebnomin  =  \frac{ \log _e(2)} 
{ E \left[ {\mathrm{Tr}} \left(\bH \bQi \bH^H \right)\right]}, \hsp 
{\rm and} \hsp  
S_0  =  2 \cdot \frac{ \left( E \left[ 
{\mathrm{Tr}} \left( \bH \bQi \bH^H  \right) \right] \right)^2  }{ 
E \left[   {\mathrm{Tr}}  \left( \left( \bH \bQi \bH^H \right)^2 
\right) \right] } 
\end{eqnarray} 
where the input covariance matrix, $\bQi = \diag \left( \bQi[i] \right)$, 
is low-$\snr$ capacity-achieving and unit trace constrained. 

When there is {\em only} one dominant transmit eigen-mode, beamforming to 
this mode is spectral efficiency-optimal. If there are $r$ dominant 
eigen-modes with $r > 1$, any $\bQi$ that excites any of the $r$ modes 
with any weightage is ergodic capacity-optimal. However,~\cite{VerduIT2002} 
points out that uniform-power signaling over these $r$ modes is necessary 
to maximize spectral efficiency. We consider these two cases separately 
in the following theorem.

\begin{theorem}
\label{speceff} 
If $r = 1$, the minimum energies per bit are given by 
\begin{eqnarray}
\sebnomin_{,\hspp \can } = \sebnomin_{, \hspp \kron } = \frac{ \log_e(2) } 
{  \sum_{i} {\bf P}_{c}[i,j_{\max}]  } \stackrel{(a)}{\rightarrow} 
\frac{\log_e(2)}{N_r p \mu } 
\label{ebnomin_case1} 
\end{eqnarray} 
where the convergence in (a) is for the sparse framework {\sf II}. An 
application of the Gaussian moment factoring theorem~\cite{gmft} with the 
optimal input shows that 
\begin{eqnarray}
S_{0,\hspp \can} & = &  2 \cdot 
\frac{ \left( \sum_{i} {\bf P}_c[i, j_{\max}] \right)^2  } 
{ \sum_{i} \left( {\bf P}_c[i,j_{\max}] \right)^2 + 
\left( \sum_{i} {\bf P}_c[i,j_{\max}] \right)^2 },  
\\ S_{0,\hspp \kron} & = & 2 \cdot 
\frac{ \left( \sum_{i} {\bf P}_k[i, j_{\max}] \right)^2  } 
{ \sum_{i} \left( {\bf P}_k[i, j_{\max} ] \right)^2 + 
\left( \sum_{i} {\bf P}_{k}[i, j_{\max}] \right)^2 }.  
\end{eqnarray} 
With framework {\sf II}, we have 
\begin{eqnarray} 
S_{0,\hspp \can} \rightarrow 
\frac{2 N_r \mu^2 p}{(N_r p + 1) \mu^2 + \sigma^2}, & {\rm and} & 
S_{0,\hspp \kron} \rightarrow \frac{2 N_r }{ N_r  + 1 }. 
\end{eqnarray} 
If $r > 1$, the energies per bit are the same as in~(\ref{ebnomin_case1}). 
The wideband slopes generalize to 
\begin{eqnarray}
S_{0, \hspp \can} \rightarrow 
\frac{2 N_r r \mu^2 p}{ \mu^2(N_r p + 1 + (r-1)p) + \sigma^2 },  
& {\rm and}& S_{0, \hspp \kron} \rightarrow 
\frac{2 N_r r}{N_r + r}. 
\end{eqnarray} 
\end{theorem}
{\vspace{0.1in}}
\begin{proof} 
With $r = 1$, the conclusion about energy per bit is straightforward. 
The expression for the wideband slope follows immediately from the fact proved 
in Appendix~\ref{app_cankron}: 
\begin{eqnarray}
N_r p^2 \mu^2 \leftarrow 
\sum_{i} \kvarijmaxfour < \sum_{i} \cvarijmaxfour \rightarrow 
N_r p \left(\mu^2 + \sigma^2 \right). 
\end{eqnarray}
For $r > 1$, see Appendix~\ref{append:app1mid}. 
\end{proof}

It can be checked that $S_{0, \hspp \can} < S_{0, \hspp \kron}$ in either case. 
However, this conclusion is not {\emph{easily}} reflected in 
Figs.~$\ref{fig_cdf_sparse}$-$\ref{fig_cdf_rich}$ due to two reasons: 
\begin{itemize} 
\item 
$\sebnomin$, 
which is the same for both the channel models (in both $r = 1$ and 
$r > 1$ cases), is the most important figure of merit at low-$\snr$ and 
corresponds to first order variation in ergodic capacity with $\snr$ while 
$S_0$ corresponds to second order variation at low-$\snr$. 
\item 
The discrepancy in $S_0$ for the two models is small. In fact, we have 
\begin{eqnarray} 
\left| S_{0, \hspp \kron} - S_{0,\hspp \can} 
\right| \leq \frac{ \sum_{i} \left(  \cvarijmaxfour - \kvarijmaxfour 
\right) } { \left( \sum_i \cvarijmax \right)^2 } = 
\frac{\sigma^2}{p\mu^2} \cdot \ord \left( \frac{1} {N_r} \right) 
\end{eqnarray}
for the $r = 1$ case, and 
\begin{eqnarray}
\left| S_{0, \hspp \kron} - S_{0,\hspp \can} \right| 
\leq \frac{\sigma^2}{p\mu^2} \cdot \ord \left( \frac{N_r r }
{(N_r + r)^2} \right) 
\end{eqnarray}
for the $r > 1$ case. In the second case, the difference in wideband slopes is 
$\ord \big( \frac{1}{N_r} \big)$ for finite values of $r$. 
\end{itemize}

\subsection{High-$\snr$ Extreme} 
We now make two assumptions on the random matrix channel $\bH_c$ to 
aid\footnote{The first condition can be relaxed with some advanced random 
matrix theory techniques that are out-of-scope here. If this is done and 
we obtain results for arbitrary $N_t$ and $N_r$, then the second condition 
can be assumed without any 
loss in generality since we can always ignore those columns$/$rows with 
zero power. Nevertheless, for simplicity of analysis, we assume both 
conditions.}  in capacity analysis: 1) $N_t = N_r = N$, and 
2) $\rank(\bH_c) = N \hsp {\mathrm{a.s.}}$ Note that the second condition 
is equivalent to assuming that none of $\{ \sum_i \cvarij \}$ and 
$\{ \sum_j \cvarij \}$ are zero. From the discussion following 
Prop.~\ref{prop_modelfit}, we 
also have $\rank(\bH_k) = N \hsp {\mathrm{a.s.}}$.

In this setting, the capacity random variables under the two models are given by 
\begin{eqnarray}
C_{\can}(\rho,\bH) \triangleq 
\log_2 \det \left( \bI_N + \frac{\rho}{N} \bH_c \bH_c^H   \right)
 \stackrel{\mathit{(a)}} { =}  \log_2 \det \left( \bH_c \bH_c^H \right) 
+ N  \log_2 \left( \frac{\rho}{N}  \right) + 
\ord \left( \frac{1}{\rho} \right) 
\label{e_capc} \\ 
C_{\kron}(\rho,\bH) \triangleq 
\log_2 \det \left( \bI_N + \frac{\rho}{N} \bH_k \bH_k^H   \right) 
\stackrel{\mathit{(b)}} { =}  \log_2 \det \left( \bH_k \bH_k^H \right) + 
N \log_2 \left( \frac{\rho}{N}  \right) 
+ \ord \left( \frac{1}{\rho} \right) 
\label{e_capk}
\end{eqnarray}
where in (a) we have used both Assumptions 1) and 2), and in (b), we have used 
the fact that $\rank\left(  \bH_k \right) = N=\rank\left(  \bH_c \right) 
\hsp {\mathrm{a.s.}}$ Hence the statistics of $C_{\can}(\rho,\bH)$ 
and $C_{\kron}(\rho,\bH)$ at high-$\snr$ are related to the 
moments of 
$\log_2 \det \left( \bH_c \bH_c^H \right)$ and $\log_2 \det \left( \bH_k 
\bH_k^H \right)$, respectively. We now perform a large-system analysis 
of these random log-determinants. 

\noindent {\bf \em Stochastic Approximation for the Canonical Case:} In 
the case of ${\bf H}_{\sf iid}$ ($\cvarij = 1$ for all $i,j$), 
this analysis is simplified by what is known as the {\emph{Bartlett 
decomposition (or bidiagonalization) of a sample covariance 
matrix}}~\cite{anderson,girko,hochwald}. The decomposition states that there 
exist independent random variables $\bZ_i$ on some probability space such 
that 
\begin{eqnarray}
\label{iid_dec}
\bZ \triangleq \det \left( \bH_{\iid} \hspp \bH_{\iid}^H \right) \sim 
\prod_{i=1}^N \bZ_i, \hsp \bZ_i \sim 
\sum_{j=i}^N \left| \bH_{\iid}[i,j] \right|^2 
\sim  \frac{1}{2} \hsp  \chi^2 \left( 2(N-i+1) \right) 
\end{eqnarray}
where $\chi^2(2k)$ is a central chi-squared random variable with $2k$ 
degrees of freedom. 

On the other hand, computing $\log_2 \det \left( \bH_c \bH_c^H \right)$ 
in closed-form is extremely difficult because $\{ \cvarij \}$, in general, 
possess no structure and a Bartlett-type decomposition for $\det \left( 
\bH_c \bH_c^H \right)$ is not known. Nevertheless, a tight stochastic 
approximation for $C_{\erg,\hspp \can }(\rho)$ is still possible and 
for this, we need the following notation from~\cite{Olkin}. 

We say that a random variable $\bX_2$ upper bounds a random variable 
$\bX_1$ (and denote it by $\bX_1 \lesssim \bX_2$) if 
\begin{eqnarray}
{\mathrm{Pr}}\left( \bX_1 < x  \right) \hsp \geq  \hsp 
{\mathrm{Pr}}\left( \bX_2 < x  \right) \hsp {\rm for} \hsp {\rm all} 
\hsp x \in {\mathbb{R}}. 
\end{eqnarray}
The following lemma provides a 
statistical ``bound'' and a useful stochastic approximation for 
$\det \left( \bH_c \bH_c^H \right)$. 
\begin{lemma} [Girko]
\label{girko_lem}
Let $\bHc[i,j]$ be independent and distributed as 
${\mathcal{CN}} (0, p_{i,j} )$. Then, 
\begin{eqnarray} 
\bZ \cdot \prod_{i=1}^N \min_j p_{i,j} 
\hsp \lesssim \hsp 
\det \left ( \bHc \bHc^H \right) 
\hsp \lesssim \hsp 
\bZ \cdot \prod_{i=1}^N \max_j p_{i,j} 
\end{eqnarray}
where $\bZ$ is as in~(\ref{iid_dec}). Moreover, there exist independent 
random variables $\bZt_i, \hsp i= 1 \hsp \cdots \hsp N$ on some probability 
space such that $\det \big ( \bHc \bHc^H \big)$ can be well-approximated 
as 
\begin{eqnarray}
\label{det_appx}
\det \big ( \bHc \bHc^H \big) \approx 
\prod_{i=1}^N \bZt_i, \hsp \bZt_i  \sim  
i \hsp \frac{ \sum_{j=1}^N \left| \bHc[i,j]  \right|^2}{N}. 
\end{eqnarray}
\end{lemma} 
{\vspace{0.1in}}
\begin{proof}
See Appendix~\ref{girko_app}. 
\end{proof}

\begin{figure}[htb!]
\centerline{\psfig{figure=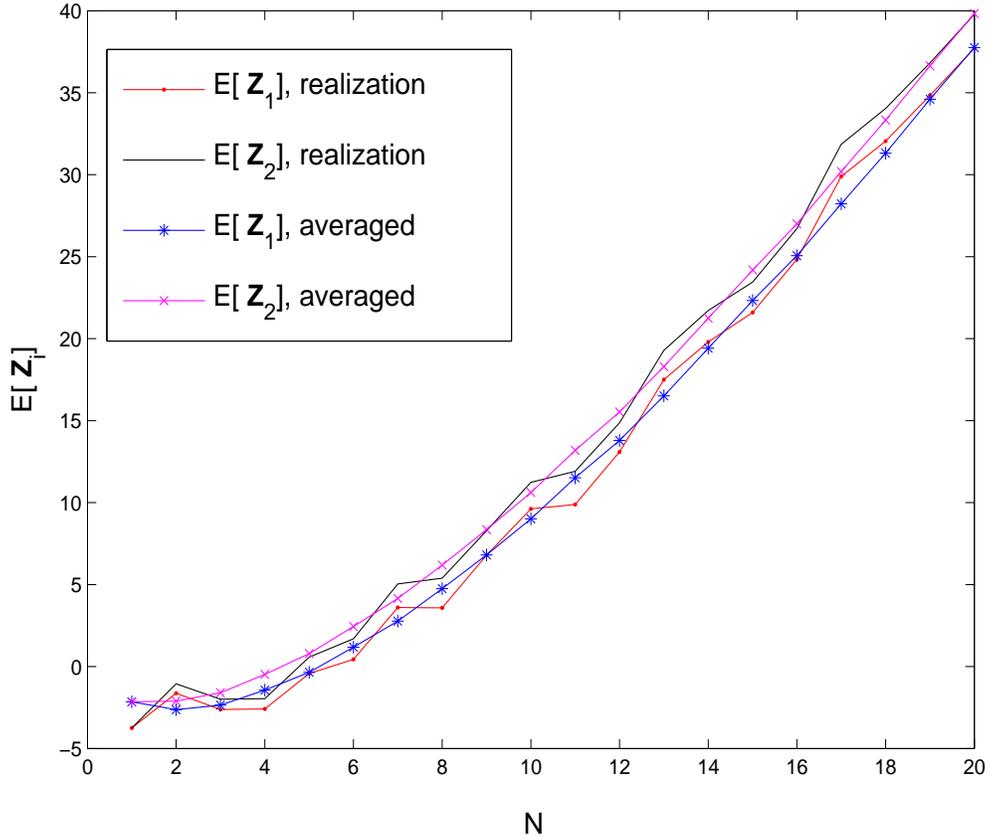,width=5.2in,height=4.5in}}
\caption{Comparison of means of ${\bf Z}_i$ (as a function of $N$) for a 
typical scattering environment and averaged over many scattering environments.}
\label{fig_randchecker}
\end{figure}

Numerical studies indicate that the approximation in Lemma~\ref{girko_lem} 
is close for a large class of random matrices {\emph{even}} for small 
values of $N$. Furthermore, this approximation gets more accurate as 
$N$ increases for a large class of random matrices. This fact is 
illustrated in~Fig.~\ref{fig_randchecker} where we plot $E[ {\bf Z}_1]$ 
and $E[ {\bf Z}_2 ]$ as a function of matrix dimension $N$ with 
\begin{eqnarray} 
{\bf Z}_1 = \log_2 \det(\bHc \bHc^H) \hsp {\rm and} \hsp 
{\bf Z}_2 = \sum_{i=1}^N \log_2 \left(  
\frac{i \hspp \sum_{j=1}^N |\bHc[i,j]|^2  }{N} \right). 
\end{eqnarray} 
The first set corresponds to a typical scattering environment where 
$\{ p_{i,j} \}$ are chosen i.i.d.\ from a uniform distribution on 
$[0,1]$ (in particular, $\mu = \frac{1}{2}$ and $\sigma^2 = \frac{1}{12}$). 
The second set corresponds to a smoothed version of the first set where 
we also average over many different scattering environments. Here, we have 
averaged over $5000$ independent scattering environments and the plot shows 
that the approximation is very accurate on average. In the rest of the paper, 
we assume that the approximation in~(\ref{det_appx}) is accurate. 
Nevertheless, its rigorous use is contingent on further studies that 
have to establish its preciseness. This will be the subject of future work.


\noindent {\bf \em Capacity Computation and Comparison:} 
\begin{theorem}
\label{mv_highsnr} 
With the sparse frameworks of Sec.~\ref{sec_sparse}, good estimates can be 
obtained for ergodic capacity in the high-$\snr$ extreme. 
\begin{itemize}
\item 
(a) The ergodic capacity under the Kronecker model converges to 
\begin{eqnarray}
{\hspace{-0.1in}} C_{\erg,\hspp \kron}(\rho) 
& \rightarrow & N \log_2 \left( \frac{ \rho \hspp N^2}
{ \sum_{ij}  p_{i,j}  } \right) + 
\sum_{i=1}^N \log_2 \bigg( \frac{i}{N} \bigg) + K_{\kron} + 
\ord \bigg( \frac{1}{\rho} \bigg) \\ 
K_{\kron} & = & \sum_{i=1}^N \log_2 \left(  
\frac{ \sum_{l} p_{i,l} \sum_{k} p_{k,i} } 
{ \sum_{kl} p_{k,l} } 
\right)
\end{eqnarray} 
whereas under the canonical model, it is well-approximated (with the 
approximation approaching an equality as $N \rightarrow \infty$ following 
the previous discussion) by 
\begin{eqnarray} 
C_{\erg, \hspp \can }(\rho) & \approx & 
N \log_2\left (  \frac{ \rho \hspp N^2} 
{ \sum_{ij} p_{i,j} } \right) 
+ \sum_{i=1}^N \log_2 \bigg( \frac{i}{N} \bigg) + 
K_{\can} + \ord \bigg( \frac{1}{\rho} \bigg)  \\ 
K_{\can}  & = & \frac{1}{2} \left\{ 
\sum_{i=1}^{N} \log_2 \bigg( 
\frac{ \sum_{j=1}^N p_{i,j} }  {N} \bigg) + 
\sum_{j=1}^{N} \log_2 \bigg( 
\frac{ \sum_{i=1}^N p_{i,j} }  {N} \bigg) \right\}. 
\label{kcan_defn}
\end{eqnarray}

\item 
(b) In the large-system regime, the following expressions are true: 
\begin{eqnarray}
C_{\erg,\hspp \can}(\rho) - C_{\erg, \hspp \kron}(\rho) \rightarrow 
\frac{N}{2} \log_2 \left( \frac{ {\rm AM}_{\sf row \hspp pow} \cdot 
{\rm AM}_{\sf col \hspp pow}  } 
{ {\rm GM}_{\sf row \hspp pow} \cdot {\rm GM}_{\sf col \hspp pow} } 
\right)
\label{am_gm_exp}
\end{eqnarray}
where ${\rm AM}_{\bullet}$ and ${\rm GM}_{\bullet}$ correspond to the 
arithmetic and geometric means of row and column powers of ${\bf P}_c$. 
Further, we also have 
\begin{eqnarray} 
\label{eqn_proved_agm}
0 \leq C_{\erg, \hspp \can}(\rho) - C_{\erg, \hspp \kron}(\rho) 
\leq 2N \log_2(N). 
\end{eqnarray} 

\item 
(c) Equality in the lower bound holds if and only if $\bH_c$ is regular. 
While it seems difficult to construct a ${\bf P}_c$ that meets the upper 
bound, the following choice is order-optimal and results in 
$C_{\erg, \hspp \can}(\rho) - C_{\erg, \hspp \kron}(\rho) 
\stackrel{N \rightarrow \infty}{\approx} N \log_2(N)$: 
\begin{eqnarray}
\label{pc_highsnr}
{\bf P}_c = \diag \Big[ N^2 - N + 1, \hspp 
\underbrace{ 1 , \hspp \cdots \hspp , 1 }_{N-1} \Big]. 
\end{eqnarray}
\end{itemize}
\end{theorem}
\begin{proof}
See Appendix~\ref{append:app3}. 
\end{proof}

\noindent {\bf \em Variance of Capacity:} Closed-form results are difficult 
to obtain for $V_{\bullet}(\rho)$ as $\rho \rightarrow \infty$. However, 
numerical studies 
indicate that for most scattering environments $\sqrt{ V_{\can}(\rho) }$ 
and $\sqrt{ V_{\kron }(\rho)}$ are sub-dominant\footnote{For example, in 
the i.i.d.\ case, it can be seen that 
$C_{\erg, \hsppp \can}(\rho) = C_{\erg, \hsppp \kron}(\rho) = \ord(N)$ while 
$V_{\can}(\rho) = V_{\kron}(\rho) = \ord(\log(N))$~\cite{hochwald}.} when 
compared with $C_{\erg, \hspp \can}(\rho)$ and $C_{\erg, \hspp \kron}(\rho)$, 
respectively. Thus, for a typical scattering environment, 
the outage capacities are primarily determined by 
$C_{\erg, \hspp \can}(\rho)$ and $C_{\erg, \hspp \kron}(\rho)$. The 
smoothing effect of the Kronecker model as can be 
seen from~(\ref{kron_var}), the low-$\snr$ trends of $V_{\bullet}(\rho)$, 
and numerical studies (see Figs.~$\ref{fig_cdf_sparse}$-$\ref{fig_cdf_rich}$) 
lend credence to the following conjecture proving which will be the subject 
of future work. 
\begin{conj}
\label{conj1}
The following are true for a large class of channels in the medium- 
to high-$\snr$ regime: 
\begin{eqnarray} 
\frac{ \sqrt{V_{\can}(\rho)} }{ C_{\erg,\hspp\can}(\rho) } 
\stackrel{ N \rightarrow \infty }{ \rightarrow } 0, \hsp 
\frac{ \sqrt{V_{\kron}(\rho)} }{ C_{\erg,\hspp\kron}(\rho) } 
\stackrel{ N \rightarrow \infty }{ \rightarrow } 0, \hsp {\rm and} \hsp 
V_{\can}(\rho) \geq V_{\kron}(\rho).  
\end{eqnarray}
\end{conj}

{\vspace{0.1in}} 
\noindent {\bf \em Discussion:} From~(\ref{am_gm_exp}), we first note that 
the mismatch accrued by the Kronecker model increases as the ratio of 
arithmetic and geometric means of the row and the column powers increases. 
The ratio of arithmetic and geometric means is a measure of the homogeneity 
of the vector (under consideration) or lack of disparities~\cite{woodhouse,jasso}: 
The regularity of ${\bf P}_c$ in our context. That is, the smaller the 
ratio, the more regular the channel and {\em vice versa}. Thus, we see 
that the more non-regular the channel, the larger the mismatch with the 
Kronecker model. This conclusion is reflected in the structure of the 
choices of ${\bf P}_c$ (in Theorem~\ref{mv_highsnr}) that lead to a large 
and a small mismatch. It is also reflected in 
Figs.~$\ref{fig_cdf_sparse}$-$\ref{fig_cdf_rich}$ where the channels become 
more regular as they become richer (This is because both the transmit and 
the receive sides become more well-conditioned as the channel becomes richer), 
and the mismatch between the Kronecker and the canonical models decreases. 

We also note the following trends. In the case of non-regular channels, 
the fact that $C_{\erg, \hspp \can}(\rho) > C_{\erg, \hspp \kron}(\rho)$ 
and the sub-dominance conjecture of $V_{\bullet}(\rho)$ implies that the 
Kronecker model underestimates capacity confirming the observations made 
in recent measurement 
campaigns~\cite{Bonek,zhou,costa_haykin,wyne,abhayapala,new_ozcelik,survey}. 
Note that the $\snr$ range of most of these observations lie between 
$10$ and $20$ dB, which can be viewed as the high-$\snr$ regime. The 
choice of the $\snr$ range also explains why the popular belief on the 
decreasing probability of overestimation~(see 
e.g.,~\cite[Footnote 5]{bonek_valid}) has come about. The case of 
regular (or near-regular) channels in the high-$\snr$ regime has a 
behavior similar to that of channels in the low-$\snr$ regime. Finally, 
note that the theory developed in this work 
is useful in the context of the probabilistic sparse framework of 
Sec.~\ref{sec_sparse} where it holds with probability $1$. Since the class 
of sparse channels forms the most predominant class in the space of all 
possible channels (Fig.~\ref{fig_sparse}), the utility of this theory is 
immense. 

\section{Conclusion}
\label{conclusion} 
In this paper, we have unified existing statistical models for spatially 
correlated multi-antenna channels by considering a canonical decomposition 
of the channel along the transmit and$/$or the receive eigen-bases. This 
framework generalizes the Kronecker model, the virtual representation 
and the Weichselberger model, and as a by-product develops two 
other classes of statistical models. In addition, we have developed an 
abstract framework to model spatial sparsity that has 
been observed in many recent measurement campaigns. 

These campaigns have also demonstrated that the Kronecker model results in 
misleading estimates for the capacity of realistic scattering environments. 
However, the reasons for these observations have not been well-understood 
so far. In this work, we have rigorously established the connection between 
spatial sparsity of the true channel, the non-regularity of the sparsity 
structure, and the impact 
they have on the capacity estimates provided by a Kronecker model fit. The 
Kronecker model fit uses the marginal sum statistics and this spreads the 
sparse DoF in the spatial domain. The consequent redistribution of the 
channel power is responsible for the mismatch in capacity estimation. In 
particular, we have shown that in the case of non-regular channels, the 
Kronecker model underestimates capacity in the medium- to high-$\snr$ regime. 
On the other hand, in the low-$\snr$ regime and regular channels in the 
high-$\snr$ regime, the Kronecker model overestimates capacity at high 
levels of operational reliability and {\em vice versa}.

Possible extensions to this work include the development of a more systematic 
framework for the generation of correlated$/$sparse multi-antenna channels, 
the impact sparsity has on the over$/$underestimation of capacity and 
reliability, establishing rigorously the approximation in 
Lemma~\ref{girko_lem} and Conjecture~\ref{conj1}, computation 
of closed-form expressions for the mean and the variance of capacity under 
the canonical and the Kronecker models at general $\snr$s, understanding 
the impact on capacity of different channel power normalizations that are 
consistent with physical intuition etc.

\appendix
\subsection{Proof of Lemma~\ref{lemmaCM3}}
\label{app0} 
Consider the matrix $\bH_c \triangleq \bU_r^H \bH \bU_t$. 
From Assumptions 1 and 2, we can write $\bH_c = \bU_r^H \bH_t$. Then, the 
cross-covariance of the columns of $\bH_c$ (denoted by $\{ \bhu_{ci} \}$) 
satisfies 
\begin{eqnarray} 
E \left[ \bhu_{ci} \hspp \bhu_{cj}^H \right] 
= \bU_r^H E \left[ \bhu_{ti} \hspp \bhu_{tj}^H\right] \bU_r 
= {\bf 0} \hsp {\rm for} \hsp {\rm all} \hsp i,j, \hsp  i \neq j, 
\end{eqnarray} 
which follows from the column uncorrelatedness of $\bH_t$ in 
Lemma~\ref{lemmaCM1}. Similarly, from Assumptions 3 and 4, we can 
write $\bH_c = \bH_r \bU_t$. Then, the cross-covariance of the rows of 
$\bH_c$ (denoted by $\{ {\bf g}_{ci} \}$) satisfies 
\begin{eqnarray} 
E\left[ {{\bf g}}_{ci} \hspp {{\bf g}}_{cj}^H \right] = \bU_t^H 
E \left[ {{\bf g}}_{ri} \hspp {{\bf g}}_{rj}^H \right] \bU_t 
= {\bf 0} \hsp {\rm for} \hsp {\rm all} \hsp i,j, \hsp  i \neq j, 
\end{eqnarray} 
which follows from the row uncorrelatedness of $\bH_r$ in 
Lemma~\ref{lemmaCM2}. Thus, the columns and the rows of $\bH_c$ are 
uncorrelated. This necessarily implies that all entries of $\bH_c$ 
are uncorrelated. 
\endproof 

\subsection{Proof of Theorem~\ref{mv_lowsnr}} 
\label{append:app1} 
\noindent {\bf \em Preliminaries}: 
The following result concerning the tail probabilities of weighted sums of i.i.d.\ 
random variables would lead us towards the estimation of $I(\rho)$. 
\begin{lemma} [Lanzinger and Stadtmueller,~\cite{lanzinger}]
\label{lem1}
Consider i.i.d.\ random variables $X, X_1, X_2, \cdots$ with $E[X] = 0, 
E[X^2] = \sigma_0^2$. Let $\beta > 0$ and $\nu \geq 2$ with 
$E[X^{\nu}] < \infty$. Define the weighted sum 
\begin{eqnarray} 
T_n \triangleq \sum_{k=1}^{n} t_k X_k, \hsp \hsp t_k \geq 0, 
\hsp \hsp {\rm and} \hsp \hsp \sigma_n^2 = \sigma_0^2 
\sum_{k=1}^n t_k^2. 
\end{eqnarray} 
Also, suppose for some $\alpha \geq 1$, 
\begin{eqnarray}
\frac{  \max \limits _{1 \leq k \leq n} t_k  } 
{\sigma_n } & \leq & \frac{\alpha}{\sqrt{n}} \hsp 
{\mathrm {for \hsp all} } \hsp n. 
\end{eqnarray}
\begin{eqnarray}
\label{refered}
{\hspace{-1.3in}}
{\mathrm{Then,}} \hsp 
\lim_{ \epsilon \rightarrow 0+ } \epsilon^{ \nu + \frac{ \nu/2 - 1}{\beta} } 
\sum_{n = 1}^{\infty} n^{ \nu(\beta + 1/2 ) - 2 } \hsp
{\mathrm{Pr}} \left( |T_n | > \epsilon n^{\beta} \sigma_n  \right) 
&= & \frac{E \big[ |\bN|^{  \nu + \frac{ \nu/2-1}{\beta} } \big]
}{\nu (\beta + 1/2) - 1}
\end{eqnarray}
where $\bN$ is a standard Gaussian random variable. 
\endproof 
\end{lemma} 
Note that the conclusion of Lemma~\ref{lem1} can be suitably modified in 
the case of $\epsilon \nrightarrow 0+$, but is sufficiently small, by 
increasing the right-hand side of (\ref{refered}) appropriately. The 
crucial point is that this would not alter our conclusion since the 
above modification can be done, by keeping the right-hand side in 
(\ref{refered}) still finite.

\noindent {\bf \em Application of Lemma~\ref{lem1}}: Lemma~\ref{lem1} 
is applied in our setting as follows. Let $X_i = 
\left| \bH_{\iid} [i, j_{\max}] \right|^2 - 1, t_i = \cvarijmax$, 
$T_n = \sum_{i=1}^n \left( \big| \bH _{c}[i, j_{\max}] \big|^2 - 
\cvarijmax \right)$, $\beta = \frac{1}{2}, \nu = 2$ and $n = N_{r}$. 
Then Lemma~\ref{lem1} implies that 
\begin{eqnarray}
\label{eqn_ref1}
\sum_{n} {\mathrm{Pr} } \left( \frac{ | T_n| } {\sum_{k=1}^n t_k } 
>  \eta \hsp\sqrt{ \frac{ \sum_{i=1}^n \cvarijmaxfour }{  n} }\cdot 
\frac{n} { \sum_{i=1}^n \cvarijmax   }  \right) \leq \frac{1}{\eta^2} < \infty
\end{eqnarray} 
for $\eta$ appropriately small. The conclusion in~(\ref{eqn_ref1}) implies 
that there exists $m > 1/2$ and $\ell > 0$ such that 
\begin{eqnarray}
{\mathrm{Pr} } \left( 
\left| \sum_{i=1}^{N_r} \big| \bH _{c}[i, j_{\max}] \big|^2 
- \cvarijmax \right| > \eta \hsp \sqrt{  
N_r \sum_{i=1}^{N_r} \cvarijmaxfour 
} \right) 
\leq \frac{1}{\eta^{2\ell} N_r^{2m}}. 
\label{prob_eqn}
\end{eqnarray} 

\noindent {\bf \em Proof of Theorem:} Let $\bY = \bY(N_r)$ denote the 
random variable $\sum_{i=1}^{N_r} \big| \bH _{c}[i, j_{\max}] \big|^2$. 
Setting $\rho = \frac{1} {\sum_{i=1}^{N_r} \bP_c[i,j_{\max}] } \cdot 
\frac{\eta}{1 - \eta} \cdot \frac{1}{\gamma}$ with 
$\gamma = 1 + \eta \cdot \frac{ \sqrt{ N_r \sum_{i=1}^{N_r} 
(\bP_c[i,j_{\max})^2] }} { \sum_{i=1}^{N_r} \bP_c[i,j_{\max}]   }$ 
and using~(\ref{prob_eqn}), we get 
\begin{eqnarray}
{\mathrm{Pr}} \left( \rho \bY > \frac{\eta}{1-\eta} \right) 
& = &  {\mathrm{Pr}} 
\left( \rho \bY - \rho \sum_{i=1}^{N_r} \bP_c[i,j_{\max}] 
> \frac{\eta}{1-\eta}  - \rho \sum_{i=1}^{N_r} \bP_c[i,j_{\max}]  
\right) 
\\ 
& \leq  & {\mathrm{Pr}} \left( \rho \left| \bY -E[\bY ] \right| > 
\eta \rho \hsp \sqrt{  N_r \sum_{i=1}^{N_r} \cvarijmaxfour } \right)
\leq \frac{1}{\eta^{2\ell} N_r^{2m}}. \nonumber 
\end{eqnarray}

An upper bound to $I(\rho)$ follows easily from the log-inequality: 
\begin{eqnarray} 
\label{upper_bd}
I(\rho) & \triangleq &  E \left[ \log_2 \left( 1 + \rho \bY \right)  
\right] 
 \leq  \rho  \log_2(e) \hsp E \left[ \bY \right] 
\hsp = \hsp  \rho \log_2(e) \sum_i \cvarijmax.  
\end{eqnarray} 
We now establish a tight lower bound for $I(\rho)$: 
\begin{eqnarray}
\label{lbd}
\frac{ E \left[ \log_2 \left( 1 + \rho \bY \right)  \right] }
{\log_2(e)} & \stackrel{\mathit{(a)}} { \geq} &  E \left[ 
\frac{ \rho \bY} { 1 + \rho \bY } \right]  \\ 
& = & \underbrace{  E \left[  \frac{ \rho \bY} { 1 + \rho \bY } \hsp 
\chi\left( \rho \bY \leq \frac{\eta}{1-\eta} \right) \right] }_{Z_1} + 
\underbrace{
E \left[  \frac{ \rho \bY} { 1 + \rho \bY } \hsp
\chi\left( \rho \bY > \frac{\eta}{1-\eta} \right) \right] }_{Z_2}
\\ & \stackrel{\mathit{(b)}} { \geq} & \underbrace{ 
E \left[ \left( \rho \bY - \rho^2 \bY^2 
\right) \hsp \chi\left( \rho \bY \leq \frac{\eta}{1-\eta} \right) \right] 
}_{Z_3} 
\\ Z_3 & = & E \left[ \rho\bY \right] - \underbrace{  
E\left[ \rho \bY \hsp 
\chi\left( \rho \bY > \frac{\eta}{1-\eta} \right) \right] }_{Z_4}
- \underbrace{ E \left[ \rho^2 \bY^2 \hsp \chi\left( \rho \bY \leq 
\frac{\eta}{1-\eta} \right)  \right]}_{Z_5} \nonumber  
\end{eqnarray}
where (a) follows from the inequality $\log_e(1+z) \geq \frac{z}{1+z}$ and (b) 
follows from using the inequality $\frac{1}{1 + z} \geq (1-z)$. An 
application of the Cauchy-Schwarz inequality shows that 
\begin{eqnarray} 
\frac{ E \left[ \log_2 \left( 1 + \rho \bY \right)  \right] }
{\log_2(e)} & \geq & \rho \cdot \sum_{i=1}^{N_r} \bP_c[i,j_{\max}] 
\underbrace{ 
\left[  1 - \frac{ \sqrt{E[\bY^2] }   } {E[\bY]} \cdot \frac{ 1} 
{\eta^{\ell} N_r^m} - \rho \cdot \frac{E[\bY^2] }{E[\bY]} \right].}_{Z_6} 
\end{eqnarray} 

Now, the quantity $Z_6$ makes meaningful sense as a lower bound to $I(\rho)$ 
only if it is positive. Plugging in the expression for $\rho$, we see that this 
can be ensured\footnote{The choice of $1/2$ for the upper bound of the valid 
interval of $\eta$ is more or less arbitrary and we have not optimized 
over this choice.} {\em if} $\eta$ is constrained to $(0, 1/2]$. Further, 
$\eta = 1/2$ maximizes the lower bound to $I(\rho)$. Evaluating $E[ \bY^2]$, 
substituting the value of $\kappa_c$ and noting that 
$\rho = \frac{ \delta }{ \sum_{i=1}^{N_r} \bP_c[i,j_{\max}] }$ for 
$\delta < \frac{1}{\gamma_0}$ where $\gamma_0$ is as in (\ref{gamma00}), we get 
\begin{eqnarray}
I(\rho) \geq \log_2(e) \cdot \delta \cdot 
\left( 1  - \frac{ 2^{\ell} \sqrt{\kappa_c} } { N_r^m } - 
\frac{\delta \kappa_c}{  \gamma_0} \right).
\end{eqnarray}
To complete the proof, we observe from~\cite{vasanth_isit07} that beamforming 
is the optimal signaling strategy for all $\rho < \rho_{\low, \hspp \can}$ 
and there exists a constant $d > 0$ (independent of $\bP_c$, $N_r$ and $N_t$) 
such that 
\begin{eqnarray}
\rho_{\low, \hspp \can } \geq \frac{d}{\sum_{i=1}^{N_r} \bP_c[i,j_{\max}]}. 
\end{eqnarray}
The constant 
$c$ in the statement of the theorem can be chosen to be $\min(d, 1/\gamma_0)$. 
It is important to note that the tightness of the upper bound in (\ref{upper_bd}) 
and the lower bound in (\ref{lbd}) critically hinge on the low-$\snr$ assumption. 
Thus the theorem is complete. 
\endproof

\subsection{Proof of Theorem~\ref{mv_cankron}}
\label{app_cankron} 
\begin{itemize} 
\item (a) From~(\ref{defn_lowsnr}), the ergodic capacities in the 
low-$\snr$ regime are given by 
\begin{eqnarray} 
C_{\erg, \hspp \can }(\rho) 
& = & E \left[ \log _2 \bigg( 1 + \rho 
\sum_{i} \big| \bH _{c}[i, j_{\max}] \big|^2 \bigg) \right], \\ 
C_{\erg, \hspp \kron }(\rho) 
& = & E \left[ \log _2 \bigg( 1 + \rho 
\sum_{i} \big| \bH _{k} [i, j_{\max}] \big|^2 \bigg) \right] 
\end{eqnarray} 
where $j_{\max} = \arg \max _j \sum_{i} \cvarij = \arg \max _j \sum_i \kvarij$. 
Estimating these quantities is a straightforward consequence of 
Theorem~\ref{mv_lowsnr}. 

\item (b) For the variance, we have 
\begin{eqnarray} 
V_{\can}(\rho) & \triangleq & 
E \left[ \left( \log_2 \left( 1 + \rho \bY \right) \right)^2  \right] - 
\left( E \left[  \log_2 \left(1 + \rho \bY \right) \right] \right)^2 
\end{eqnarray}
where $\bY = \sum_{i=1}^{N_r} \big| \bH _{c}[i, j_{\max}] \big|^2$. 
Proceeding along similar lines as in App.~\ref{append:app1}, we have 
\begin{eqnarray}
V_{\can }(\rho) & \leq & 
\left( \log_2(e) \right)^2 \cdot \delta^2 \cdot 
\left( \kappa_c - \left( 1  - 
\frac{ 2^{\ell} \sqrt{\kappa_c} } { N_r^m } -  
\frac{\delta \kappa_c}{  \gamma_0} 
\right)^2 \right) \\ 
V_{\can }(\rho) 
& \geq & \left( \log_2(e) \right)^2 \cdot \delta^2 \cdot 
\left( \kappa_c -1 - \frac{ \sqrt{E[\bY^4] } \cdot 2^{\ell} } 
{ \left( \sum_{i=1}^{N_r} \bP_c[i,j_{\max}] \right)^2 \cdot N_r^m} 
- \frac{2 \delta \cdot E[\bY^3]} 
{ \left( \sum_{i=1}^{N_r} \bP_c[i,j_{\max}] \right)^3 } \right) 
\nonumber 
\end{eqnarray}
where the constants are as in the statement of Theorem~\ref{mv_lowsnr}. 
Thus, we can recast $V_{\can }(\rho)$ as 
\begin{eqnarray}
V_{\can  }(\rho) = \left( \log_2(e) \right)^2  \cdot \rho^2 \cdot 
\sum_{i} \cvarijmaxfour \cdot \left( 1 + \littleo(1) \right) 
\label{vcrho}
\end{eqnarray} 
where the $\littleo(1)$ factor in the above expression converges to $0$ 
as $N_r \rightarrow \infty$ and $\rho \rightarrow 0$. The critical 
assumption in the above proof is that $\bH_c[i,j]$ are independent 
random variables. Thus, the same proof technique can be adapted to 
compute $V_{\kron}(\rho)$ as well.

\item (c) 
The relationship between $V_{\can}(\rho)$ and $V_{\kron}(\rho)$ is 
not obvious. For this, we need the following result on the monotonicity 
of ratios of means\cite[pp.\ 129-130]{Olkin}. 
\begin{lemma} [Marshall, Olkin and Proschan] 
\label{lem_majorizer}
Let ${\bf x} = [x_1, \cdots, x_n]$ and ${\bf y} = [y_1, \cdots, y_n]$ 
be two vectors such that $\sum_{i=1}^n x_i = \sum_{i = 1}^n y_i$. If 
$\frac{y_i}{x_i}$ is decreasing in $i$ and $x_1 \geq \cdots \geq x_n > 0$, 
then ${\bf x}$ is majorized by ${\bf y}$, and 
\begin{eqnarray}
g(r) \triangleq \left( \frac{ \sum_{i=1}^n x_i^r  } { \sum_{i=1}^n y_i^r } 
\right)^{1/r}
\end{eqnarray} 
is decreasing in $r$ for $r > 0$. \endproof 
\end{lemma} 

\noindent {\bf \em Application of Lemma~\ref{lem_majorizer}:} We set 
\begin{eqnarray}
x_i \triangleq \frac{ \sum_{k=1}^{N_t} {\bf P}_c[i,k] }{\rho_c}, 
& & 
y_i \triangleq \frac{ {\bf P}_c[i, j _{\max}] } 
{  \sum_{k=1}^{N_r} {\bf P}_c[k,j_{\max}]  } 
\end{eqnarray} 
and $n = N_r$. From the assumption in the statement of the theorem, note 
that $x_i > 0$ for all $i$ and are in decreasing order. The fact that 
$\frac{y_i}{x_i}$ is decreasing is a consequence of (\ref{ass_yixi}). A 
straightforward consequence of Lemma~\ref{lem_majorizer} is that 
$V_{\can}(\rho) \geq V_{\kron}(\rho)$. \endproof 
\end{itemize}

\subsection{Proof of Prop.~\ref{prop_as}} 
\label{app_prop_as} 

\begin{itemize}
\item 
(a) With framework {\sf II}, the main goal is to compute the probability 
of failure of~(\ref{ass_yixi}). Towards this computation, we first 
condition upon $p_{i,j_{\max} }$ and $p_{i+1, j_{\max}}$ (in particular, 
$s_{\bullet}$ and $q_{\bullet}$) where $i$ is such that $1 \leq i \leq N_r-1$. 
Define the conditional probability $p_i$: 
\begin{eqnarray}
p_i & \triangleq & {\rm Pr} \left( 
\frac{ \sum_{k \neq j_{\max} } {\bf P}_c[i,k]  }
{ \sum_{k \neq j_{\max} } {\bf P}_c[i+1, k] } > 
\frac{ {\bf P}_c[i, j_{\max}] }{ {\bf P}_c[i+1,j_{\max}] } \right) 
\\ & = & {\rm Pr} \left( 
p_{i+1, j_{\max} } \sum_{k \neq j_{\max}} q_{i,k} s_{i,k}  - 
p_{i, j_{\max} } \sum_{k \neq j_{\max} } q_{i+1,k} s_{i+1,k}   
> 0 \right). 
\end{eqnarray}
Hence, the conditional probability of failure of~(\ref{ass_yixi}) is 
$1 - \prod_{i=1}^{N_r-1} (1 - p_i)$. We intend to show that the above 
probability converges to $0$ as $N_t \rightarrow \infty$. 

Without loss in generality, we can assume that $s_{i,j_{\max}} = 
s_{i+1,j_{\max}} = 1$ (Otherwise, $p_i = 0$.). Similarly, we can 
assume that $\{ q_{i,j_{\max}} \}$ is decreasing in $i$. Note that 
$p_i$ can be written as 
\begin{eqnarray}
p_i & = & {\rm Pr} \left( \frac{ {\bf Z}_{N_t}}{N_t } > 0 \right) 
\hsp \hsp {\rm where}
\\ {\bf Z}_{N_t} & = &  q_{i+1, j_{\max} }  
\sum_{k \neq j_{\max}} q_{i,k} s_{i,k} 
- q_{i, j_{\max} }  \sum_{k \neq j_{\max}}  q_{i+1,k} s_{i+1,k} . 
\end{eqnarray}
Using the independence of $\{ q_{i,k} \}$ and $\{ s_{i,k} \}$ 
and their statistics, it can be checked that 
\begin{eqnarray}
\frac{ E \left[ {\bf Z}_{N_t} \right] } { N_t } 
& \stackrel{N_t \rightarrow \infty}{\rightarrow} & 
\left( q_{i+1, j_{\max} } - q_{i, j_{\max} } \right) p \mu < 0, \\ 
{\rm Var} \left( \frac{ {\bf Z}_{N_t}}{N_t}  \right) 
& \stackrel{N_t \rightarrow \infty} {\rightarrow} & 
\frac{ \left( \left( q_{i+1, j_{\max} }\right)^2 + 
\left(q_{i,j_{\max} } \right)^2 \right) p \sigma^2}{N_t} \rightarrow 0. 
\end{eqnarray} 
That is, $\frac{ {\bf Z}_{N_t} }{N_t}$ hardens around its mean (which is 
negative) as $N_t \rightarrow \infty$ and hence, $p_i$ converges to $0$. 
Averaging over $\{ p_{i,j_{\max} } \}$, we see that for ``almost all'' 
sparse scattering environments, the condition in~(\ref{ass_yixi}) holds 
and hence, $V_{\can}(\rho) \geq V_{\kron}(\rho)$.

\item (b) We now compare the dominant terms of the variances of capacity 
with the two models. We have 
\begin{eqnarray}
\frac{ V_{\can  }(\rho) } 
{ N_r \hspp \left( \log_2(e) \rho \right)^2  } = 
\frac{ \sum_i p_{i,j_{\max}}^2 }{N_r} 
\cdot \frac{1} {  \left( \frac{ \sum_{kl} p_{k,l} } {N_t N_r} \right)^2  } .
\end{eqnarray} 
After using~(\ref{kron_var}), we can also write $V_{\kron}(\rho)$ in 
terms of $\{ p_{i,j} \}$ as 
\begin{eqnarray}
\frac{ V_{\kron}(\rho) } 
{ N_r \hspp \left( \log_2(e) \rho \right)^2  } 
& = & 
\frac{  \left( \frac{\sum_k p_{k, j_{\max}}  }{N_r} \right)^2
\cdot \frac{1}{N_r} \sum_i  \left( \frac{ \sum_{l} p_{i,l} }{N_t} 
\right)^2  } 
{ \left( \frac{ \sum_{kl} p_{k,l} }{N_t N_r} \right)^4   }. 
\end{eqnarray} 

In the large-system regime, since $\{ p_{i,j} = q_{i,j} s_{i,j} \}$ and 
$\{ q_{i,j} \}$ is a realization from an i.i.d.\ family of mean~$\mu$ 
and variance~$\sigma^2$, we can use the law of large 
numbers~\cite{durrett} to check that 
\begin{eqnarray}
\frac{ V_{\can }(\rho) } 
{ N_r \hspp \left( \log_2(e) \rho \right)^2  } 
\rightarrow \frac{ E[ (q_{i,j} )^2  ] } 
{ \left( E[q_{i,j}] \right)^2 p  }, 
\; \; \; \; \;  
\frac{ V_{\kron }(\rho) } 
{ N_r \hspp \left( \log_2(e) \rho \right)^2  } 
\rightarrow 1.  
\label{almostsure}
\end{eqnarray} 

The fact that $V_{\can }(\rho) \geq V_{\kron }(\rho)$ follows from the 
Cauchy-Schwarz inequality. The upper bound for 
$\frac{ V_{\sf can}(\rho) }{V_{\sf kron}(\rho)}$ follows from the reverse 
Cauchy-Schwarz inequality~\cite[equation 24, p.\ 208]{bullen} due to Cassels, 
which is stated here for convenience. 
\begin{lemma}
\label{reverse_cassels} 
If ${\bf x} = [x_1, \cdots, x_n]$, ${\bf y} = [y_1, \cdots , y_n]$ 
and ${\bf w} = [w_1, \cdots, w_n]$ are 
positive $n$-tuples such that $0 < m_1 \leq x_i \leq M_1$ and 
$0 < m_2 \leq y_i \leq M_2$ for all $i$ with $m_1 m_2 < M_1 M_2$, then 
\begin{eqnarray}
\frac{  \left( \sum_{i=1}^n x_i^2 w_i^2 \right) \cdot 
\left( \sum_{i=1}^n y_i^2 w_i^2 \right) }
{ \left( \sum_{i=1}^n x_i y_i w_i^2 \right)^2 } \leq 
\frac{  \left( m_1 m_2 + M_1 M_2 \right)^2 }{ 4 m_1 m_2 M_1 M_2 }. 
\end{eqnarray}
\endproof 
\end{lemma} 

\item 
(c) Equality in the lower bound is possible if and only if $q_{i,j}$ is 
constant with probability $1$ and $p = 1$. 
That is, ${\bf H}_c$ and ${\bf H}_k$ are 
i.i.d. With ${\bf P}_c$ as in~(\ref{pc_ubound}), it can be checked that 
\begin{eqnarray}
\frac{V_{\sf can}(\rho)  }{ V_{\sf kron}(\rho) } = 
\frac{ \left( M + m \right)^2 }{4 Mm } \cdot 
\frac{ 1 + \frac{ 2m + M }{m N} } { \left( 1 + \frac{M+m}{2m N} 
\right)^2 } \stackrel{N \rightarrow \infty}{ \rightarrow } 
\frac{ \left( M + m \right)^2 }{4 Mm }. 
\end{eqnarray}
\end{itemize}
Thus the proposition is complete. 
\endproof

\subsection{Proof of Theorem~\ref{speceff}}
\label{append:app1mid} 
When $r > 1$, we assume that the $r$ dominant columns have been relabeled as 
columns $1$ through $r$. We then have 
\begin{eqnarray}
\sebnomin_{,\hspp \can } = \sebnomin_{, \hspp \kron } = \frac{r \log_e(2) } 
{  \sum_{i=1}^{N_r} \sum_{j=1}^{r} \cvarij  } = 
\frac{ \log_e(2) } 
{  \sum_{i=1}^{N_r} \cvarijmax  } 
\end{eqnarray} 
with the last equality following because all the $r$ columns have the same 
sums. Using the uniform input over $r$ modes and the Gaussian moment factoring 
theorem, the wideband slopes can be checked to be  
\begin{eqnarray}
S_{0,\hspp \can} & = &  2 \cdot 
\frac{ \left( \sum_{i=1}^{N_r} \sum_{j=1}^{r} \cvarij \right)^2  } 
{ \sum_{i=1}^{N_r} \sum_{j_1 = 1}^{r} \sum_{j_2 = 1}^{r} \bP_c[i,j_1] \bP_c[i,j_2] 
+ \sum_{j=1}^r \left(\sum_{i=1}^{N_r} \bP_c[i,j] \right)^2  },  
\\ S_{0,\hspp \kron} & = & 2 \cdot 
\frac{ \left( \sum_{i=1}^{N_r} \sum_{j=1}^{r} \kvarij \right)^2  } 
{ \sum_{i=1}^{N_r} \sum_{j_1 = 1}^{r} \sum_{j_2 = 1}^{r} \bP_k[i,j_1] \bP_k[i,j_2] 
+ \sum_{j=1}^r \left(\sum_{i=1}^{N_r} \bP_k[i,j] \right)^2  }. 
\end{eqnarray}
Using the law of large numbers appropriately, we have 
\begin{eqnarray}
S_{0, \hspp \can} \rightarrow 
\frac{2 N_r r \mu^2 p }{ \mu^2( N_r p + (r-1)p + 1) + \sigma^2 }, 
&{\rm and}& S_{0, \hspp \kron} \rightarrow 
\frac{2 N_r r}{N_r + r}. 
\end{eqnarray} 
\endproof

\subsection{Proof of Lemma~\ref{girko_lem}} 
\label{girko_app} 
See \cite[Chap.\ 2, p.\ 104]{girko} for a proof of the first statement. 
For the statement on determinant approximation, we start with  
\cite[p.\ 35, 39]{girko_book2} which states that $\det \big( 
\bHc \bHc^H \big)$ can be decomposed as a product of independent 
random variables, $\bZh_i$ where 
\begin{eqnarray} 
\bZh_i \sim \sum_{j=1}^{i} \left| \eta_{ij} 
\right|^2, \hsp \hsp \eta_{ij} = \sum_{l=1}^N \bHc[i,l] 
\hsp\theta[l,j], 
\end{eqnarray} 
and the matrix $\Theta = \{ \theta[i,j] \}$ is a unitary 
random matrix independent of $\bHc$. Note that 
\begin{eqnarray}
\left| \eta_{ij} \right|^2 = \sum_l \left|\bHc[i,l] \right|^2 \hsp 
\left|\theta[l,j] \right|^2 + \sum_{ l_1 \neq l_2} 
\bHc[i,l_1] \bHc[i,l_2]^{\star} \theta[l_1,j] \theta[l_2,j]^{\star}
\end{eqnarray}
and using the facts that the entries of a random unitary matrix are 
asymptotically self-averaging, (that is, zero mean in a ``statistical''  
sense) and the rows and columns have unit norm, we have the following 
approximation for $\left| \eta_{ij} \right|^2$:  
\begin{eqnarray}
\left| \eta_{ij} \right|^2 \approx \frac{1}{N} 
\sum_l \left|\bHc[i,l] \right|^2. 
\end{eqnarray} 
This leads to the approximation for $\bZh_i$, which we denote by 
$\bZt_i$ in the statement of lemma. 
\endproof

\subsection{Proof of Theorem~\ref{mv_highsnr}}
\label{append:app3} 
\begin{itemize} 
\item (a) The i.i.d.\ result can be exploited in the Kronecker case as 
follows: 
\begin{eqnarray}
\label{abc}
\log_2 \det \left( \bH_k \hspp \bH_k ^H \right) & 
\stackrel{\mathit{(a)}}{= } & \log_2 \det \left( 
\bLambda_r \hspp \bH_{\iid} \hspp \bLambda_t \hspp \bH_{\iid}^H  \right) 
\\ 
& \stackrel{(b)}{=} & \log_2\det( \bLambda_t \hspp \bLambda_r ) + \log_2 \det 
\left( \bH_{\iid} \hspp \bH_{\iid}^H \right ) \\ 
& \stackrel{\mathit{(c)}}{\sim} & 
\sum_{i=1}^N \log_2 \left(  \frac{1}{2} \cdot \kvarii 
\cdot \chi^2 \left( 2(N-i+1) \right)  \right) 
\end{eqnarray}
where (a) follows from the definition of $\bH_k$, (b) from the fact 
that $N_t = N_r$ and 
$\det({\bf AB}) = \det({\bf BA})$, and (c) from (\ref{iid_dec}) 
and the definitions of $\bLambda_t$ and $\bLambda_r$. Using 
$E \left[\log_2 \det(\bH_{\iid} \bH_{\iid}^H )\right]$~\cite{hochwald}, 
we can compute 
$C_{\erg, \hspp \kron }(\rho)$ 
to be 
\begin{eqnarray} 
C_{\erg,\hspp \kron }(\rho) & \rightarrow & 
N \log_2 \left( \frac{ \rho \hspp N^2} { \sum_{ij} p_{i,j} } \right ) 
+  \sum_{i=1}^N \log_2 \bigg( \frac{i}{N} \bigg) + K_{\kron} 
+ \ord \bigg( \frac{1}{\rho} \bigg) 
\end{eqnarray}
where 
\begin{eqnarray}
K_{\kron} & = & 
\sum_{i=1}^N \log_2 \left(  
\frac{ \sum_{l} p_{i,l} \sum_{k} p_{k,i}  } 
{ \sum_{kl} p_{k,l} } \right). 
\end{eqnarray}

For the canonical case, we write ${\bf H}_c {\bf H}_c^H$ as 
\begin{eqnarray}
{\bf H}_c {\bf H}_c^H = \frac{N^2 }{ \sum_{ij}p_{i,j} }  
\bHc \bHc^H, \hspp \bHc[i,j] \sim {\cal CN}(0, p_{i,j})
\end{eqnarray}
and compute $C_{\erg, \hspp \can}(\rho)$ as follows: 
\begin{eqnarray}
\label{ty1} 
E \left[ \log_2 \det ( \bHc \bHc^H) \right] 
& \stackrel{ {\mathit{(a)}} } {  \approx } & 
\sum_{i=1}^N \log_2(i) + 
\sum_{i=1}^{N} E \left[ \log_2 \bigg( \frac{ \sum_{j=1}^N 
| \bHc[i,j] |^2 }{N} \bigg) \right] \\ 
& \stackrel{ {\mathit{(b)}} } {  \rightarrow } & 
\sum_{i=1}^N \log_2(i) +  \underbrace{ \sum_{i=1}^{N} \log_2 \bigg( 
\frac{ \sum_{j=1}^N p_{i,j}   }{N}  \bigg) }_{K_{\can}} 
\\ 
C_{\erg,\hspp \can }(\rho ) & \approx & 
N \log_2\left ( \frac{ \rho \hspp N^2} 
{ \sum_{ij} p_{i,j} } \right) + \sum_{i=1}^N 
\log_2 \bigg( \frac{i}{N} \bigg)  + K_{\can} + 
\ord \bigg( \frac{1}{\rho} \bigg)  \nonumber 
\end{eqnarray} 
where (a) follows from the approximation (the approximation gets more 
accurate as $N \rightarrow \infty$) in Lemma~\ref{girko_lem}. 
The convergence in (b) follows 
from Prop.~\ref{prop_unifintegrable} which is stated and proved next. 
Since $N_r = N_t = N$, all the above steps are true even if ${\bf H}_c$ 
is replaced with ${\bf H}_c^H$. This leads to the expression for $K_{\can}$ 
in~(\ref{kcan_defn}). 

\begin{prop}
\label{prop_unifintegrable} 
With the setting as above, we have 
\begin{eqnarray} 
\log_2 \bigg( \frac{ \sum_{j=1}^N | \bHc[i,j] |^2 }{N} \bigg) 
\stackrel{N \rightarrow \infty}{\rightarrow} 
\log_2 \bigg( \frac{ \sum_{j=1}^N p_{i,j}   }{N}  \bigg) 
\hspp \hspp {\rm in} \hspp {\rm mean} \hspp {\rm for} \hspp {\rm any} 
\hspp i. 
\end{eqnarray} 
\end{prop}
{\vspace{0.08in}}
\begin{proof} 
We decompose the left-hand side as 
\begin{eqnarray} 
&& E \left[ \log_2 \left( \frac{ \sum_{j=1}^N | \bHc[i,j] |^2 }{N} \right) 
\right] \nonumber 
\\ & = &  E \left[ \log \left( \frac{ \sum_{j = 1}^N 
|\bHc[i,j]|^2 \chi \left(  |\bHc[i,j]|^2 \leq K \right)  
+ |\bHc[i,j]|^2 \chi \left( |\bHc[i,j]|^2 > K \right)} {N} \right)
\right] \nonumber 
\\ & 
= & E \left[ \log \left( \frac{ \sum_{j=1}^N |\bHc[i,j]|^2 
\chi \left( |\bHc[i,j]|^2 \leq K \right) } {N} 
\right) \right] \nonumber \\ 
& {\hspace{0.3in}} + & 
E \left[ \log \left( 1 + 
\frac{ \sum_{j=1}^N |\bHc[i,j]|^2 \chi \left( |\bHc[i,j]|^2 > K \right) } 
{\sum_{j=1}^N |\bHc[i,j]|^2 \chi \left( |\bHc[i,j]|^2 \leq K \right)}
\right) \right] 
\end{eqnarray}
for some $K > 0$ fixed. 

For the first term, note that the weak law of large numbers states 
that for all $i$ 
\begin{eqnarray}
\frac{ \sum_{j=1}^N | \bHc[i,j] |^2 \chi \left(  | \bHc[i,j] |^2 \leq K 
\right)}{N} & \stackrel{N \rightarrow \infty}{\rightarrow} & 
\frac{ \sum_{j=1}^N E \left[ | \bHc[i,j] |^2 
\chi \left(  | \bHc[i,j] |^2 \leq K \right) \right]  }{N} 
\\ & = & {P}, \nonumber \\ 
P & \triangleq & \frac{ \sum_{j = 1}^N p_{i,j} - 
\left( p_{i,j} + K \right)  e^{-\frac{K }{ p_{i,j}  }} }{N}. 
\end{eqnarray} 
The convergence is in probability and hence, also 
weakly~\cite[p.\ 310]{grimmett}. The second equality follows from a 
routine expectation computation. Since $\log(\cdot)$ is a continuous 
function and the limit random variable is a constant, 
following~\cite[p.\ 316, p.\ 310]{grimmett} we also have 
\begin{eqnarray} 
\log \left( \frac{ \sum_{j=1}^N | \bHc[i,j] |^2  \chi \left( 
|\bHc[i,j]|^2 \leq K \right) }{N} \right) \stackrel{p} 
{\rightarrow} \log \left( P \right).  
\end{eqnarray} 
The above convergence can further be strengthened to convergence in 
mean since the random variables are bounded by $K$ for all $i$ and 
all choices of $N$~\cite[p.\ 310]{grimmett}. 

For the second term, we use the following lower bound: 
\begin{eqnarray}
|\bHc[i,j]|^2 \chi \left( |\bHc[i,j]|^2 \leq K \right) 
\geq 
|\bHc[i,j]|^2 \chi \left( \epsilon < |\bHc[i,j]|^2 \leq K  \right) 
\end{eqnarray}
for some $0 < \epsilon \leq K$. Using this, we can upper bound the 
second term by 
\begin{eqnarray} 
& {\hspace{-0.5in}} &
 E \left[ \log \left( 1 + 
\frac{ 
\sum_{j=1}^N |\bHc[i,j]|^2 \chi \left( |\bHc[i,j]|^2 > K \right) } 
{\sum_{j=1}^N |\bHc[i,j]|^2 
\chi \left( \epsilon < |\bHc[i,j]|^2 \leq K \right)}
\right) \right] \nonumber \\ & {\hspace{0.2in}} \leq & 
E \left[ \log \left( 1 + \frac{ \sum_{j=1}^N |\bHc[i,j]|^2 
\chi \left( |\bHc[i,j]|^2 > K \right) } {N \epsilon} \right) 
\right] 
\\ & {\hspace{0.2in}} \leq &  
\frac{ \sum_{j=1}^N E \left[   
|\bHc[i,j]|^2 \chi \left( |\bHc[i,j]|^2 > K \right) \right]  }{N \epsilon} 
\\ 
& {\hspace{0.2in}} =& 
\frac{ \sum_{j = 1}^N \left( p_{i,j} + K \right) 
e^{- \frac{K}{ p_{i,j} }  } } {N\epsilon} 
\end{eqnarray} 
where the second step follows from the log-inequality. Combining these 
two results by choosing $K$ sufficiently large to ensure that 
$\left( p_{i,j} + K \right) e^{- \frac{K}{ p_{i,j}  } }$ is 
sufficiently small for all $i, \hspp j$ and $\epsilon$ finite, we 
obtain the conclusion as in the statement of the proposition. 
\end{proof} 

\item 
(b) In the large-system regime, we have 
\begin{eqnarray}
C_{\erg, \hspp \can}(\rho) - C_{\erg, \hspp \kron}(\rho) \approx 
K_{\can} - K_{\kron} 
& = & \sum_{i=1}^N \log_2 \left( 
\frac{ \sum_{kl}p_{k,l} } {N \sqrt{ \sum_{l} p_{i,l} \sum_{k} p_{k,i}  } } 
\right) \\ 
& = & 
\log_2 \left( \frac{N^N}{  \left( \prod_{i=1}^N P_i Q_i \right)^{1/2} }   
\right) \\ 
& = & \frac{N}{2} \log_2 \left( \frac{ 
{\rm AM}_{\sf row \hspp pow } \cdot {\rm AM}_{ \sf col \hspp pow } } 
{ {\rm GM}_{ \sf row \hspp pow }  \cdot {\rm GM}_{ \sf col \hspp pow }  }
\right)
\end{eqnarray}
where $P_i = \sum_{j= 1}^N {\bf P}_c[j,i]$ and 
$Q_i = \sum_{j=1}^N {\bf P}_c[i,j]$ are the column and the row powers, 
respectively such that $\sum_{i} P_i = \sum_i Q_i = N^2$. 

An application of the arithmetic-geometric mean inequality shows that 
$K_{\can} \geq K_{\kron}$. For an upper bound on the difference, we use 
the reverse arithmetic-geometric mean 
inequality~\cite[Theorem 3, p.\ 124]{bullen} due to Docev, which is stated 
here for convenience. 
\begin{lemma}
\label{reverse_agm}
If ${\bf x} = [x_1, \cdots, x_n]$ is a positive $n$-tuple with $K = 
\frac{\max_i x _i}{\min_i x_i}$, then 
\begin{eqnarray} 
\frac{ {\rm AM}_{ {\bf x} } } { {\rm GM}_{ {\bf x} } } \leq 
\frac{(K-1) K^{\frac{1}{K-1} } } 
{ e \log(K) } . 
\end{eqnarray}
\endproof 
\end{lemma} 
Since ${\bf P}_c$ is $\rank$-$N$, we apply Lemma~\ref{reverse_agm} with 
$K = N^2 - N +1$ for an upper bound, and the result is~(\ref{eqn_proved_agm}). 

\item 
(c) Equality in the application of the arithmetic-geometric mean inequality is 
possible if and only if $P_i = Q_i = N$ for all $i$. It is straightforward 
to check that a channel satisfying this property has to be necessarily 
regular (see Footnote~\ref{footnote_regular}). The 
conclusion for the lower bound follows by plugging the choice of ${\bf P}_c$ 
in~(\ref{pc_highsnr}) in the capacity expressions. 
\end{itemize}
\endproof

\bibliographystyle{IEEEbib}
\bibliography{newrefs,refs}


\end{document}

%% file: macros.tex
\def\convergewp1{{\> {\buildrel {w.p.1} \over \longrightarrow} \> }}

\newcommand{\Reals}{\mathop{\hbox{\mit I\kern-.2em R}}\nolimits}

\newcommand{\bn}{{\bf n}}

\newcommand{\bHc}{ {\widetilde{{\bf H}}} }

\newcommand{\bec}{\begin{center}}
\newcommand{\ec}{\end{center}}
\newcommand{\barr}{\begin{array}}
\newcommand{\earr}{\end{array}}
\newcommand{\bx}{{\bf x}}

\newcommand{\bh}{{\bf h}}
\newcommand{\by}{{\bf y}}

\newcommand{\Nats}{\mathop{\hbox{\mit I\kern-.2em N}}\nolimits}

\newcommand{\beq}{\begin{equation}}
\newcommand{\eeq}{\end{equation}}
\newcommand{\beqr}{\begin{eqnarray}}
\newcommand{\eeqr}{\end{eqnarray}}
\newcommand{\bcr}{\begin{center}}
\newcommand{\ecr}{\end{center}}

\newtheorem{theorem}{Theorem}

\newtheorem{lemma}{Lemma}
\newtheorem{prop}{Proposition}
\newtheorem{conj}{Conjecture}

\newcommand{\Sigmabf}{{\mbox{\boldmath $\Sigma$}}}

\newcommand{\bLambda}{ {\bf \Lambda} }

\newcommand{\rank}{{\sf rank } }
\newcommand\ignore[1]{}
\newcommand{\kron} {  {\sf kron} } 
\newcommand{\can}{ {\sf can} }

\newcommand{\CM}{ {\sf CM3} }
\newcommand{\CMone}{ {\sf CM1} } 
\newcommand{\CMtwo}{ {\sf CM2} }
\newcommand{\punnormij}{ p_{i,j} }

\newcommand{\cvarij}{{\mathbf{P}}_{c}[i,j]}
\newcommand{\kvarij}{{\mathbf{P}}_{k}[i,j]}

\newcommand{\kvarii}{{\mathbf{P}}_{k}[i]}

\newcommand{\kvarijmaxfour } {  \big( {\mathbf{P}}_{k }[i,j_{\max}] \big)^2}
\newcommand{\cvarijmaxfour } { \big( {\mathbf{P}}_{c 
}[i,j_{\max}] \big)^2 }
\newcommand{\cvarijmax} {{\mathbf{P}}_{c }[i,j_{\max}]}

\newcommand{\bR}{{\bf R}}
\newcommand{\bU}{{\bf U}}
\newcommand{\bQi}{  {\bf Q}}

\newcommand{\bI}{{\bf I}}

\newcommand{\bZt} {  \widetilde{ {\mathbf{Z}} } }
\newcommand{\bZh} {  \widehat{ {\mathbf{Z}} } }
\newcommand{\bZ} { {\mathbf{Z}} }
\newcommand{\bX} { {\mathbf{X}} }
\newcommand{\bH}{ {\mathbf{H}}  }
\newcommand{\bY} { {\mathbf{Y}} }
\newcommand{\bN} { {\mathcal{N}} }
\newcommand{\bA}{{\mathbf{A}}}

\newcommand{\DoF} {  {\mathrm{DoF}} }

\newcommand{\hsp}{\hspace{0.05in} }
\newcommand{\hspp}{\hspace{0.03in} }
\newcommand{\hsppp}{\hspace{0.01in}}
\newcommand{\snr}{{\sf{SNR}} }
\newcommand{\ord}{{\mathcal{O}}}
\newcommand{\littleo}{{\mathnormal{o}}}
\newcommand{\sebnomin} { {\mathnormal{  {\frac{E_b } {N_0}_{ {\mathrm{min}} }}  } }}
\newcommand{\opt} {{\sf{opt}}  }

\newcommand{\erg} {{\sf{erg}}  }
\newcommand{\out} {{\sf{out}}  }

\newcommand{\iid} {  {\sf{iid}} }

\newcommand{\compnorm} { {\mathcal{CN}} }
\newcommand{\low} {  {\sf{low}} }
\newcommand{\chanpow}{ \rho_c  }
\newcommand{\high} {  {\sf{high}} }

\newcommand{\diag} {  {\sf{diag}} }

\newcommand{\rich} {  {\sf{rich}} }
\newcommand{\sparse} {  {\sf{sparse}} }

%% file: cap_jrs9.bbl
\begin{thebibliography}{10}

\bibitem{Telatar}
\'{I}.~E. Telatar,
\newblock ``{Capacity of Multi-Antenna Gaussian Channels},''
\newblock {\em Eur. Trans. Telecommun.}, vol. 10, pp. 585--596, Nov. 1999.

\bibitem{Foschini}
G.~J. Foschini,
\newblock ``{Layered Space-Time Architechture for Wireless Communication in a
  Fading Environment when Using Multi-Element Antennas},''
\newblock {\em Bell Labs Tech. J.}, vol. 1, no. 2, pp. 41--59, 1996.

\bibitem{Chuah}
C-N. Chuah, J.~M. Kahn, and D.~N.~C. Tse,
\newblock ``{Capacity Scaling in MIMO Wireless Systems under Correlated
  Fading},''
\newblock {\em IEEE Trans. Inform. Theory}, vol. 48, no. 3, pp. 637--650, Mar.
  2002.

\bibitem{Shiu}
D-S. Shiu, G.~J. Foschini, M.~Gans, and J.~M. Kahn,
\newblock ``{Fading Correlation and Its Effect on the Capacity of Multielement
  Antenna Systems},''
\newblock {\em IEEE Trans Commun.}, vol. 48, no. 3, pp. 502--513, Mar. 2000.

\bibitem{review_bol}
D.~Gesbert, H.~Bolcskei, D.~A. Gore, and A.~J. Paulraj,
\newblock ``{Outdoor MIMO Wireless Channels: Models and Performance
  Prediction},''
\newblock {\em IEEE Trans. Commun.}, vol. 50, no. 12, pp. 1926--1934, Dec.
  2002.

\bibitem{Sayeeddecon}
A.~M. Sayeed,
\newblock ``{Deconstructing Multi-Antenna Fading Channels},''
\newblock {\em IEEE Trans. Sig. Proc.}, vol. 50, no. 10, pp. 2563--2579, Oct.
  2002.

\bibitem{spl_issue}
K.~Liu, V.~Raghavan, and A.~M. Sayeed,
\newblock ``{Capacity Scaling and Spectral Efficiency in Wideband Correlated
  MIMO Channels},''
\newblock {\em IEEE Trans. Inform. Theory}, vol. 49, no. 10, pp. 2504--2526,
  Oct. 2003.

\bibitem{Veeravallicap}
V.~V. Veeravalli, Y.~Liang, and A.~M. Sayeed,
\newblock ``{Correlated MIMO Rayleigh Fading Channels: Capacity, Optimal
  Signaling and Asymptotics},''
\newblock {\em IEEE Trans. Inform. Theory}, vol. 51, no. 6, pp. 2058--2072,
  June 2005.

\bibitem{ada}
A.~S.~Y. Poon, R.~W. Broderson, and D.~N.~C. Tse,
\newblock ``{Degrees of Freedom in Multiple-Antenna Channels: A Signal Space
  Approach},''
\newblock {\em IEEE Trans. Inform. Theory}, vol. 51, no. 2, pp. 523--536, Feb.
  2005.

\bibitem{Kai}
K.~Yu, M.~Bengtsson, B.~Ottersten, P.~Karlsson, and M.~A. Beach,
\newblock ``{Second Order Statistics of NLOS Indoor MIMO Channels Based on 5.2
  GHz Measurements},''
\newblock {\em IEEE Global Telecommun. Conf.}, vol. 1, pp. 156--160, Nov. 2001.

\bibitem{mcnamara}
D.~P. McNamara, M.~A. Beach, and P.~N. Fletcher,
\newblock ``{Spatial Correlation in Indoor MIMO Channels},''
\newblock {\em IEEE Intern. Symp. Pers. Ind. Mob. Radio Commun.}, vol. 1, pp.
  290--294, Sept. 2002.

\bibitem{kron_int1}
J.~Kermoal, L.~Schumacher, K.~Pedersen, P.~Mogensen, and F.~Frederiksen,
\newblock ``{A Stochastic MIMO Radio Channel Model with Experimental
  Validation},''
\newblock {\em IEEE Journ. Sel. Areas in Commun.}, vol. 20, no. 6, pp.
  1211--1226, Aug. 2002.

\bibitem{wallace}
J.~W. Wallace, M.~A. Jensen, A.~L. Swindlehurst, and B.~D. Jeffs,
\newblock ``{Experimental Characterization of the MIMO Wireless Channel: Data
  Acquisition and Analysis},''
\newblock {\em IEEE Trans. Wireless Commun.}, vol. 2, no. 2, pp. 335--343, Mar.
  2003.

\bibitem{mimo_manhattan}
D.~Chizhik, J.~Ling, P.~W. Wolniansky, R.~A. Valenzuela, N.~Costa, and
  K.~Huber,
\newblock ``{Multiple-Input-Multiple-Output Measurements and Modeling in
  Manhattan},''
\newblock {\em IEEE Journ. Sel. Areas in Commun.}, vol. 21, no. 3, pp.
  321--331, Apr. 2003.

\bibitem{KotechachannelestICC}
J.~H. Kotecha and A.~M. Sayeed,
\newblock ``{Optimal Signal Design for Estimation of Correlated MIMO
  Channels},''
\newblock {\em IEEE Intern. Conf. Commun.}, vol. 5, pp. 3170--3174, May 2003.

\bibitem{Kotechachannelest}
J.~H. Kotecha and A.~M. Sayeed,
\newblock ``{Transmit Signal Design for Optimal Estimation of Correlated MIMO
  Channels},''
\newblock {\em IEEE Trans. Sig. Proc.}, vol. 52, no. 2, pp. 546--557, Feb.
  2004.

\bibitem{Kotechacaptech}
J.~H. Kotecha and A.~M. Sayeed,
\newblock ``{Canonical Statistical Models for Correlated MIMO Fading Channels
  and Capacity Analysis},''
\newblock {\em Technical Report \#ECE-03-05, University of Wisconsin-Madison},
  Mar. 2004,
\newblock Available: [Online]. {\tt{http://dune.ece.wisc.edu/}}.

\bibitem{Bonek}
W.~Weichselberger, M.~Herdin, H.~\"{O}zcelik, and E.~Bonek,
\newblock ``{A Stochastic MIMO Channel Model with Joint Correlation of Both
  Link Ends},''
\newblock {\em IEEE Trans. Wireless Commun.}, vol. 5, no. 1, pp. 90--100, Jan.
  2006.

\bibitem{tulino_ind}
A.~M. Tulino, A.~Lozano, and S.~Verd\'{u},
\newblock ``{Impact of Antenna Correlation on the Capacity of Multiantenna
  Channels},''
\newblock {\em IEEE Trans. Inform. Theory}, vol. 51, no. 7, pp. 2491--2509,
  July 2005.

\bibitem{zhou}
Y.~Zhou, M.~Herdin, A.~M. Sayeed, and E.~Bonek,
\newblock ``{Experimental Study of MIMO Channel Statistics and Capacity via the
  Virtual Channel Representation},''
\newblock {\em Technical Report, University of Wisconsin-Madison}, Feb. 2007,
\newblock Available: [Online]. {\tt{http://dune.ece.wisc.edu/}}.

\bibitem{costa_haykin}
N.~Costa and S.~Haykin,
\newblock ``{A Novel Wideband Channel Model and Experimental Validation},''
\newblock {\em IEEE Trans. Antennas and Propagat.}, vol. 56, no. 2, pp.
  550--562, Feb. 2008.

\bibitem{new_ozcelik}
H.~Ozcelik, N.~Czink, and E.~Bonek,
\newblock ``{What Makes a Good MIMO Channel Model?},''
\newblock {\em IEEE Spring Veh. Tech. Conf.}, vol. 1, pp. 156--160, May 2005.

\bibitem{wyne}
S.~Wyne, A.~Molisch, P.~Almers, G.~Eriksson, J.~Karedal, and F.~Tufvesson,
\newblock ``{Statistical Evaluation of Outdoor-to-Indoor Office MIMO
  Measurements at 5.2 GHz},''
\newblock {\em IEEE Fall Veh. Tech. Conf.}, vol. 1, pp. 146--150, May 2005.

\bibitem{abhayapala}
T.~A. Lamahewa, R.~A. Kennedy, T.~D. Abhayapala, and T.~Betlehem,
\newblock ``{MIMO Channel Correlation in General Scattering Environments},''
\newblock {\em Aus. Commun. Theory Workshop}, pp. 93--98, Feb. 2006.

\bibitem{survey}
P.~Almers, E.~Bonek, and A.~Burr et~al.,
\newblock ``{Survey of Channel and Radio Propagation Models for Wireless MIMO
  Systems},''
\newblock {\em EURASIP Journ. Wireless Commun. and Networking}, vol. 2007,
  2007.

\bibitem{bonek_valid}
E.~Bonek,
\newblock ``{Experimental Validation of Analytical MIMO Channel Models},''
\newblock {\em Elektrotechnik und Informationstechnik (e\&i)}, vol. 122, no. 6,
  pp. 196--205, 2005.

\bibitem{Tulino}
A.~M. Tulino, A.~Lozano, and S.~Verd\'{u},
\newblock ``{Capacity-Achieving Input Covariance for Correlated Multi-Antenna
  Channels},''
\newblock {\em Allerton Conf. Commun. Cont. and Comp.}, 2003.

\bibitem{wood_hodgkiss}
L.~Wood and W.~S. Hodgkiss,
\newblock ``{A Reduced-Rank Eigenbasis MIMO Channel Model},''
\newblock {\em IEEE Wireless Telecommun. Symp.}, pp. 78--83, Apr. 2008.

\bibitem{bucci}
O.~M. Bucci and G.~Franceschetti,
\newblock ``{On Degrees of Freedom of Scattered Fields},''
\newblock {\em IEEE Trans. Antennas and Propagat.}, vol. 37, no. 7, pp.
  918--926, July 1989.

\bibitem{migliore}
M.~D. Migliore,
\newblock ``{On the Role of the Number of Degrees of Freedom of the Field in
  MIMO Channels},''
\newblock {\em IEEE Trans. Antennas and Propagat.}, vol. 54, no. 2, pp.
  620--628, Feb. 2006.

\bibitem{massimo}
K.~Chakraborty and M.~Franceschetti,
\newblock ``{Maxwell Meets Shannon: Space-Time Duality in Multiple Antenna
  Channels},''
\newblock {\em Allerton Conf. Commun. Cont. and Comp.}, 2006.

\bibitem{xu_janaswamy}
J.~Xu and R.~Janaswamy,
\newblock ``{Electromagnetic Degrees of Freedom in 2-D Scattering
  Environments},''
\newblock {\em IEEE Trans. Antennas and Propagat.}, vol. 54, no. 12, pp.
  3882--3894, Dec. 2006.

\bibitem{hanlen_sparse}
L.~W. Hanlen, R.~Timo, and R.~Perera,
\newblock ``{On Dimensionality for Sparse Multipath},''
\newblock {\em Aus. Commun. Theory Workshop}, pp. 125--129, Feb. 2006.

\bibitem{rod_kennedy_sparse}
H.~M. Jones, R.~A. Kennedy, and T.~D. Abhayapala,
\newblock ``{On Dimensionality of Multipath Fields: Spatial Extent and
  Richness},''
\newblock {\em IEEE Intern. Conf. Acoustics, Speech, Sig. Proc.}, vol. 3, pp.
  2837--2840, May 2002.

\bibitem{goodman}
N.~A. Goodman,
\newblock ``{MIMO Channel Rank via the Aperture-Bandwidth Product},''
\newblock {\em IEEE Trans. Wireless Commun.}, vol. 6, no. 6, pp. 2246--2254,
  June 2007.

\bibitem{it_rs01}
V.~Raghavan and A.~M. Sayeed,
\newblock ``{Weak Convergence and Rate of Convergence of MIMO Capacity Random
  Variable},''
\newblock {\em IEEE Trans. Inform. Theory}, vol. 52, no. 8, pp. 3799--3809,
  Aug. 2006.

\bibitem{vasanth_limbf}
V.~Raghavan, R.~W. Heath, Jr., and A.~M. Sayeed,
\newblock ``{Systematic Codebook Designs for Quantized Beamforming in
  Correlated MIMO Channels},''
\newblock {\em IEEE Journ. Sel. Areas in Commun.}, vol. 25, no. 7, pp.
  1298--1310, Sept. 2007.

\bibitem{vasanth_limprecode}
V.~Raghavan, V.~V. Veeravalli, and A.~M. Sayeed,
\newblock ``{Quantized Multimode Precoding in Spatially Correlated MIMO
  Channels},''
\newblock {\em Submitted to IEEE Trans. Sig. Proc.}, 2008,
\newblock Available: [Online]. {\tt{http://arxiv.org/abs/0801.3526}}.

\bibitem{vas_matched}
V.~Raghavan, A.~M. Sayeed, and V.~V. Veeravalli,
\newblock ``{Low-Complexity Structured Precoding for Spatially Correlated MIMO
  Channels},''
\newblock {\em Submitted to IEEE Trans. Inform. Theory}, 2008,
\newblock Available: [Online]. {\tt{http://arxiv.org/abs/0805.4425}}.

\bibitem{noncoh_mahesh}
S.~G. Srinivasan and M.~K. Varanasi,
\newblock ``{Constellation Design for the Noncoherent MIMO Rayleigh Fading
  Channel at General SNR},''
\newblock {\em IEEE Trans. Inform. Theory}, vol. 53, no. 4, pp. 1572--1584,
  Apr. 2007.

\bibitem{noncoh_mahesh2}
S.~G. Srinivasan and M.~K. Varanasi,
\newblock ``{Optimal Constellations for the Low SNR Noncoherent MIMO Fading
  Channel},''
\newblock {\em Submitted to IEEE Trans. Inform. Theory}, 2007.

\bibitem{Kotechaglobecomm2003}
J.~H. Kotecha, Z.~Hong, and A.~M. Sayeed,
\newblock ``{Coding and Diversity Gain Tradeoff in Space-Time Codes for
  Correlated MIMO Channels},''
\newblock {\em IEEE Global Telecommun. Conf.}, vol. 2, pp. 646--650, Dec. 2003.

\bibitem{vasanth_clin_vvv}
C.~Lin, V.~Raghavan, and V.~V. Veeravalli,
\newblock ``{To Code or Not to Code Across Time: Space-Time Coding with
  Feedback},''
\newblock {\em To appear, IEEE Journ. Sel. Areas in Commun.}, Oct. 2008.

\bibitem{girko}
V.~L. Girko,
\newblock {\em {Theory of Random Determinants}},
\newblock Kluwer, MA, 1990.

\bibitem{Visotsky}
E.~Visotsky and U.~Madhow,
\newblock ``{Space-Time Transmit Precoding with Imperfect Feedback},''
\newblock {\em IEEE Trans. Inform. Theory}, vol. 47, no. 6, pp. 2632--2639,
  Sept. 2001.

\bibitem{liang}
Y.~Liang and V.~V. Veeravalli,
\newblock ``{Correlated MIMO Rayleigh Fading Channels: Capacity and Optimal
  Signaling},''
\newblock {\em Proc. IEEE Asilomar Conf. Signals, Systems and Computers}, vol.
  1, pp. 1166--1170, Nov. 2003.

\bibitem{vasanth_isit07}
V.~Raghavan, V.~V. Veeravalli, and R.~W. Heath, Jr.,
\newblock ``{Reduced Rank Signaling in Spatially Correlated MIMO Channels},''
\newblock {\em IEEE Intern. Symp. Inform. Theory}, 2007.

\bibitem{wyner}
L.~H. Ozarow, S.~Shamai (Shitz), and A.~Wyner,
\newblock ``{Information Theoretic Considerations for Cellular Mobile Radio},''
\newblock {\em IEEE Trans. Veh. Tech.}, vol. 43, no. 2, pp. 359--378, May 1994.

\bibitem{Goldsmith}
A.~J. Goldsmith, S.~A. Jafar, N.~Jindal, and S.~Vishwanath,
\newblock ``{Capacity Limits of MIMO Channels},''
\newblock {\em IEEE Journ. Sel. Areas in Commun.}, vol. 21, no. 5, pp.
  684--702, June 2003.

\bibitem{it_rs06}
V.~Raghavan and A.~M. Sayeed,
\newblock ``{Multi-Antenna Capacity of Sparse Multipath Channels},''
\newblock {\em Submitted to IEEE Trans. Inform. Theory}, 2008,
\newblock Available: [Online]. {\tt{http://www.ifp.uiuc.edu/$\sim$vasanth}}.

\bibitem{VerduIT2002}
S.~Verd\'{u},
\newblock ``{Spectral Efficiency in the Wideband Regime},''
\newblock {\em IEEE Trans. Inform. Theory}, vol. 48, no. 6, pp. 1319--1343,
  June 2002.

\bibitem{gmft}
I.~S. Reed,
\newblock ``{On a Moment Theorem for Complex Gaussian Processes},''
\newblock {\em IRE Trans. Inform. Theory}, vol. 8, pp. 194--195, Apr. 1962.

\bibitem{anderson}
T.~W. Anderson,
\newblock {\em {An Introduction to Multivariate Statistical Analysis}},
\newblock John Wiley, NY, 1st edition, 1960.

\bibitem{hochwald}
B.~M. Hochwald, T.~L. Marzetta, and V.~Tarokh,
\newblock ``{Multiple-Antenna Channel Hardening and its Implications for Rate
  Feedback and Scheduling},''
\newblock {\em IEEE Trans. Inform. Theory}, vol. 50, no. 9, pp. 1893--1909,
  Sept. 2004.

\bibitem{Olkin}
A.~W. Marshall and I.~Olkin,
\newblock {\em {Inequalities: Theory of Majorization and its applications}},
\newblock Academic Press, NY, 1979.

\bibitem{woodhouse}
I.~H. Woodhouse,
\newblock ``{The Ratio of the Arithmetic to the Geometric Mean: A Cross-Entropy
  Interpretation},''
\newblock {\em IEEE Trans. Geoscience and Remote Sensing}, vol. 39, no. 1, pp.
  188--189, Jan. 2001.

\bibitem{jasso}
G.~Jasso,
\newblock ``{Measuring Inequality: Using the Geometric Mean/Arithmetic Mean
  Ratio},''
\newblock {\em Sociological Methods \& Research}, vol. 10, no. 3, pp. 303--326,
  Feb. 1982.

\bibitem{lanzinger}
H.~Lanzinger and U.~Stadtmueller,
\newblock ``{Refined Baum-Katz Laws for Weighted Sums of I.I.D. Random
  Variables},''
\newblock {\em Stat. and Prob. Lett.}, vol. 69, pp. 357--368, 2004.

\bibitem{durrett}
R.~A. Durrett,
\newblock {\em {Probability: Theory and Examples}},
\newblock Duxbury Press, 2nd edition, 1995.

\bibitem{bullen}
P.~S. Bullen, D.~S. Mitrinovic, and P.~M. Vasic,
\newblock {\em {Means and Their Inequalities}},
\newblock D. Reidel Publishing Company, 1988.

\bibitem{girko_book2}
V.~L. Girko,
\newblock {\em {Theory of Linear Algebraic Equations with Random
  Coefficients}},
\newblock Allerton Press, NY, 1996.

\bibitem{grimmett}
G.~Grimmett and D.~Stirzaker,
\newblock {\em Probability and Random Processes},
\newblock Oxford, UK, 3rd edition, 2001.

\end{thebibliography}
